\documentclass[12pt]{article}
\usepackage{latexsym}
\usepackage{epsfig,graphics}

\newcommand{\be}{\begin{equation}}
\newcommand{\ee}{\end{equation}}

\def\lsi{\raise0.3ex\hbox{$<$\kern-0.75em\raise-1.1ex\hbox{$\sim$}}}
\def\gsi{\raise0.3ex\hbox{$>$\kern-0.75em\raise-1.1ex\hbox{$\sim$}}}

\begin{document}

\begin{titlepage}

\null\vspace{-1.0cm}

\begin{tabbing}
\` Oxford OUTP-01-17P \\
\end{tabbing}

\begin{centering} 

\null\vspace{1.5cm}

{\large \bf   SU(N) gauge theories in four dimensions :  \\
exploring the approach to ${\bf N = \infty}$ }

\vspace{1.5cm}
B. Lucini and M. Teper

\vspace{0.9cm}
{\it Theoretical Physics, University of Oxford, 1 Keble Road, \\
Oxford OX1 3NP, UK\\}

\vspace{2.25cm}
{\bf Abstract.}
\end{centering}

\noindent
We calculate the string tension, $\sigma$, and some of the lightest 
glueball masses, $m_G$, in 3+1 dimensional SU($N$) lattice gauge 
theories for $2 \leq N \leq 5$. From the continuum extrapolation of the
lattice values, we find that the mass ratios $m_G/\sqrt\sigma$ 
appear to show a rapid approach to the large--$N$ limit, and, indeed, 
can be described all the way down to SU(2) using just a leading 
$O(1/N^2)$ correction. We confirm that the smooth large--$N$ limit
we find, is obtained by keeping a constant 't Hooft coupling. 
We also calculate the topological charge of the gauge fields.
We observe that, as expected, the density of small-size instantons 
vanishes rapidly as $N$ increases, while the topological susceptibility 
appears to have a non-zero $N = \infty$ limit.

\vfill

\end{titlepage}

\section{Introduction}
\label{sec_intro}

How SU($N$) gauge theories approach their $N=\infty$
limit, and what that limit is, is an interesting question
\cite{largeN}
whose answer would represent a significant step towards 
addressing the same question in the context of QCD.
Accurate lattice calculations in 2+1 dimensions 
\cite{mt98}
show that in that case the approach is remarkably precocious 
in that even $N=2$ is close to $N=\infty$. Such calculations 
have to be very accurate, of course, because for each 
value of $N$ one has to perform a continuum extrapolation
of various mass ratios and then these are compared and 
extrapolated to $N=\infty$. Existing D=3+1 calculations 
\cite{mtoldN,MTrev98,winoh}
are much too rough for this purpose even if their message
appears to be optimistic. 

In this paper we present a calculation in 3+1 dimensions 
that is accurate enough for some conclusions to be drawn.
This calculation is intended as an exploratory one,
designed to see if a much more detailed and extensive
calculation is warranted. For this reason we have limited
ourselves to calculating just three masses in the 
(`glueball') spectrum: the lightest and first excited
$J^{PC}=0^{++}$ scalars, and the lightest $2^{++}$ tensor.
We have also calculated the topological susceptibility and 
have obtained some information on the distribution of
`instanton' sizes so as to see how it evolves at larger $N$.
We simultaneously investigate linear confinement, calculating
the string tension. We do all this for SU(4) and SU(5)
gauge theories and compare the results to what one finds
for SU(2) and SU(3). In fact, because the existing SU(2)
calculations proved too inaccurate to be useful, we found that
we had to redo SU(2) as well. We have also chosen to perform 
our own SU(3)
calculations (although this was not absolutely necessary)
so that when we compare results at various $N$ in this 
paper, they will have all been obtained with exactly the 
same methods.

For $N > 3$ there are, in addition, new stable strings
that connect sources in representations higher than the
fundamental and we have calculated their string tensions. In
\cite{blmtstring1}
we presented our preliminary results on this topic. We have 
since somewhat increased the accuracy of those calculations
as well as performing some similar, but much more accurate, 
calculations in D=2+1. We do not include these calculations
in the present paper since they do not really belong to
the question being addressed here, i.e. the approach of
SU($N$) gauge theories to their $N=\infty$ limit.
Instead we will present them elsewhere
\cite{blmtstring2},
as this  will  enable us to present in some detail the
theoretical background to the ideas involved. Here we
simply remark that our previous conclusions
\cite{blmtstring1}
remain unchanged: in the  D=3+1  SU(4) and SU(5) 
gauge theories the doubly
charged string has a tension that is much less than twice
that of the fundamental string, and it agrees, within fairly 
small errors, with the M(-theory)QCD conjecture made in
\cite{MQCD}.
However, as we show in
\cite{blmtstring2}
our results also agree with old speculations about the
`Casimir scaling' of strings, a dependence that is manifested 
in leading-order Hamiltonian strong-coupling and which 
happens to be numerically quite close to the MQCD conjecture.
This aspect was not discussed in
\cite{blmtstring1}
but is explored in detail in
\cite{blmtstring2}.
Interest in Casimir scaling has recently revived 
\cite{DeldarCS,BaliCS,SimCS}
following recent studies of unstable higher representation 
strings in SU(3)
\cite{DeldarCS,BaliCS}.
In our much more accurate D=2+1 calculations it is clear that 
these new stable string tensions, while being close to the MQCD 
conjecture, do in fact deviate from it. What we see
is closer to Casimir scaling although here too the
agreement does not appear to be perfect. (Although there
is some sign that the deviations
disappear quite rapidly with increasing $N$.) The fact that
the strings we deal with are stable removes ambiguities in
interpreting string tensions of the unstable  SU(3) strings
and brings into sharp focus the possible
relevance of Casimir scaling to the dynamics of confinement
\cite{blmtstring2}.

In the next section we summarise the technical details
of our lattice calculations. We then study the strong-to-weak
coupling transition in the case of SU(4). The practical
reason for doing this is to demonstrate that the range of
lattice spacings we shall use for our continuum
extrapolation avoids this potential phase transition.
However the nature of this transition
is of interest in itself and it becomes more interesting
as $N$ increases. We discuss this in some detail.
In Section \ref{sec_spectrum} we present
our results on glueball masses and the string tension
in SU(2), SU(3), SU(4) and SU(5) gauge theories.
We perform the continuum extrapolation of various
mass ratios, compare their $N$-dependence
and perform an extrapolation to $N=\infty$. We
discuss what needs to be done better in future calculations. 
In Section \ref{sec_coupling} we ask whether our
non-perturbative calculations support the usual
diagrammatic expectation 
\cite{largeN}
that a smooth large-$N$ limit is obtained by keeping
the 't Hooft coupling $\lambda \equiv g^2 N$ constant.
We then turn to calculations of the topological 
susceptibility and of the size distribution of the
topological charges, with a view to clarifying the
fate of topological fluctuations as $N$ increases.
We finish with some concluding remarks.

We remark that a quite similar calculation of the physical 
properties of D=3+1 SU($N$) gauge theories
is being carried out simultaneously elsewhere
\cite{lddevN}.
\section{Lattice preliminaries}
\label{sec_prelim}

Our four dimensional lattice is hypercubic and has periodic 
boundary conditions. The degrees of freedom are SU($N$) 
matrices, $U_l$, residing on the links, $l$, of the lattice.
In the partition function the fields are weighted with
$\exp\{S\}$ where $S$ is the standard plaquette action
\be
S = - \beta \sum_p \biggl (1 - {1\over N} ReTr \, U_p \biggr ),
\label{eqn_action}
\ee
i.e. $U_p$ is the ordered product of the matrices on
the boundary of the plaquette $p$. For smooth fields this
action reduces to the usual continuum action with
$\beta = 2N/g^2$. By varying the inverse
lattice coupling $\beta$ we vary the lattice
spacing $a$.  

Our Monte Carlo mixes standard heat-bath and over-relaxation 
steps in the ratio $1:4$. These are implemented by updating
SU(2) subgroups using the Cabibbo-Marinari prescription
\cite{cabmar}.
We use 3 subgroups in the case of SU(3), 6 for SU(4)
and 10 for SU(5). As a check of efficient ergodicity
we use the same algorithm to minimise the action  and
we find that with this number of subgroups the SU($N$) 
lattice action does decrease more-or-less as effectively
as in the SU(2) gauge theory.

Our typical lattice calculation, at a given value of $\beta$
and for a given volume, involves $10^5$ Monte Carlo sweeps
through the lattice. (For some of the coarser lattice
spacings, where the correlation functions decrease most rapidly,  
we perform more sweeps.) We perform calculations of correlation
functions every 5'th such sweep. Calculations of the topological
charge (whose details we leave to Section \ref{subsec_latticeQ}) 
are performed every 50 sweeps. 

We calculate correlations of gauge-invariant operators $\phi(t)$, 
which depend on field variables within a given time-slice, $t$.
The basic component of such an operator will typically be the
(traced) ordered product of the $U_l$ matrices around some closed
contour $c$. A contractible contour, such as the plaquette itself,
is used for glueball operators. A non-contractible closed contour, 
which winds around the spatial hyper-torus, projects onto winding
strings of fundamental flux. In the confining phase the theory is
invariant under a class of centre transformations that ensure
that the overlap between such contractible and non-contractible  
operators is exactly zero.
For our lattice action the correlation function of such an operator 
has good positivity properties, i.e. we can write
\be
C(t) = \langle \phi^{\dagger}(t) \phi(0) \rangle
= 
\sum_n | \langle \Omega | \phi | n \rangle |^2
\exp \{- E_n t \}
\label{eqn_corrln}
\ee
where $|n\rangle$ are the energy eigenstates, with $E_n$ the 
corresponding energies, and $|\Omega\rangle$ is the vacuum state. 
If the operator has $\langle \phi\rangle = 0$ then the vacuum will
not contribute to this sum and we can extract the mass of the
lightest state with the quantum numbers of $\phi$, from the large-$t$
exponential decay of $C(t)$. To make the mass calculation more 
efficient we use ${\vec p} = 0$ operators. Note that on a lattice
of lattice spacing $a$ we will have
$t=a n_t$, where $n_t$ is an integer labelling the time-slices,
so that what we actually obtain from eqn(\ref{eqn_corrln}) is  
$aE_n$, the energy in lattice units.

In practice a calculation using the simplest lattice string 
operator is inefficient because the overlap onto the lightest 
string state is small and so one has to go to large values of 
$t$ before the contribution of excited states has died away;
and there the signal disappears into the statistical noise.
There are standard methods
\cite{blocking}
for curing this problem, using blocked (smeared) link operators 
and variational techniques. This is described in detail in, 
for example, 
\cite{mt98}.
In this exploratory study  we use the simplest blocking technique 
\cite{mt98}
to produce blocked link matrices, and we form operators by multiplying
these blocked links around $1\times 1$ and $1\times 2$ contours.
We take linear combinations which transform according to the 
cubic $A^{++}_1$ and $E^{++}$ representations, and this allows us
to extract $J^{PC} = 0^{++}$ and $J^{PC} = 2^{++}$ masses.
To make the calculation more efficient we use a variational
criterion that determines the linear combination of our operators
that has the best overlap onto the lightest  $0^{++}$ and $2^{++}$
glueball states, and onto the first excited  $0^{++}$ state. These
standard techniques are described, for example, in
\cite{mt98}.

For any given state we determine the best operator as described above.
We then attempt to fit the corresponding correlation function, 
normalised so that $C(t=0) = 1$, with a single exponential in $t$. 
(Actually a $\cosh$ to take into account the
temporal periodicity.) We choose fitting intervals $[t_1,t_2]$
where initially $t_1$ is chosen to be $t_1=0$ and then is increased
until an acceptable fit is achieved. The value of  $t_2$ is chosen
so that there are at least 3, and preferably 4, values of $t$
being fitted. (Since our fitting function has two parameters.)
Where $t_1 = 0$ and the errors on $C(t=a)$ are much smaller
than the errors at $t\geq 2a$, this procedure provides no significant
evidence for the validity of the exponential fit, and so we use the 
much larger error from  $C(t=2a)$ rather than  $C(t=a)$. (This 
typically arises on the coarsest lattices and/or for very massive 
states.) We ignore correlations between statistical errors
at different $t$ and attempt to compensate for this both by demanding 
a lower $\chi^2$ for the best acceptable fit and by not extending
unnecessarily the fitting range. (Although in practice the error
on the best fit increases as we increase the fitting range,
presumably because the correlation in $t$ of the errors is modest 
and the decorrelation of the operator correlations is less efficient
as $t$ increases.) The relatively rough temporal 
discretisation of many of our calculations, means that, at the
margins, there are inevitable ambiguities in this procedure.
These however decrease as $a\to 0$. Once a fitting range is chosen, 
the error on the mass is obtained by a Jack-knife procedure which 
deals correctly with any error correlations as long as the binned 
data are statistically independent. Typically we take 50 bins,
each involving some 2000 sweeps. It is plausible that bins of
this size are independent; however we have not stored our
results in a sufficiently differential form that we can
calculate the autocorrelation functions so as to test this in 
detail. A crude test is provided by recalculating 
the statistical errors using bins that are twice as large.
We find the errors are essentially unchanged, 
which provides some evidence for the statistical
independence of our original bins.

\section{The strong--to--weak coupling transition}
\label{sec_transition}

It has long been known that if we use the standard plaquette
action then we find a cross-over between
the strong and weak coupling regions, which is characterised
by an anomalous dip in the mass of the lightest scalar
glueball and a peak in the specific heat. This effect
becomes more marked as we go from SU(2) to  SU(3) and
it is believed that in the case of SU($N\geq 4$) one has
an actual bulk phase transition in this cross-over region.
We want to locate the crossover to make sure
that it does not interfere with our continuum extrapolations.
We shall do so for the case of SU(4) where there have been
rough estimates 
\cite{mtoldN}
that it is located in the region $\beta \in [10.3,10.5]$.

\begin	{figure}[p]
\begin	{center}
\epsfig{figure=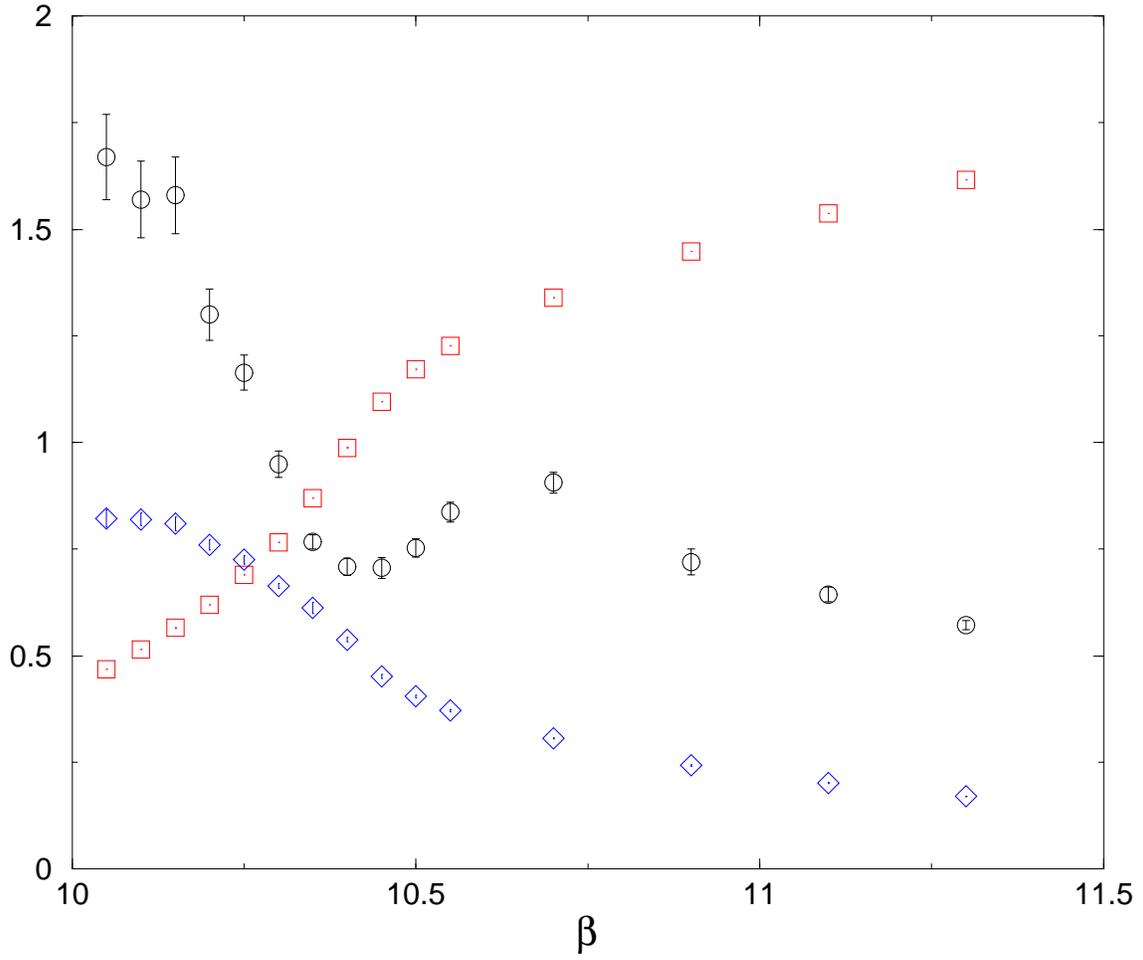, angle=270, width=15cm} 
\end	{center}
\vskip 0.15in
\caption{The (rescaled) average plaquette, $\Box$, the mass 
gap, $\circ$, and the square root of the string tension,
$\Diamond$, over a range of $\beta$ that includes the 
region of transition between
strong and weak coupling, for the SU(4) gauge theory.}
\label{fig_transition}
\end 	{figure}

To locate the cross-over region we calculate the lightest
scalar glueball mass, the string tension  and the plaquette, 
in the range of couplings $\beta \in [10.0,11.3]$. Since
we think of the lattice spacing as increasing with $\beta$ 
we would naively expect the value of $am$, for any mass $m$,
to decrease uniformly over this range. And indeed, as we see
in Fig.\ref{fig_transition}, this is the case for the
square root of the string tension, $a\surd\sigma$.
However, as we see in Fig.\ref{fig_transition}, this
is not the case for the lightest scalar glueball,
$am_{0^{++}}$, which has a striking dip at
$\beta \sim 10.4$. We note that in this region the 
value of $ds/d\beta$, where $s$ is the average plaquette,
has a maximum, which tells us that there will be a peak
in the specific heat here. Taking into account that the
dip in $am_{0^{++}}$ is superimposed on a montonically
decreasing function, we estimate that the crossover
is centered somewhere in  $\beta \in [10.35,10.40]$.
The calculations we shall use to extrapolate to the
continuum limit will all be well to the weak-coupling
side of this value of $\beta$.
 
As far as the presence of an actual phase transition is
concerned, we see no sign of a strong first order
transition in that there appear to be no discontinuities in
any of the quantities we calculate. We have not done
the kind of finite volume study that might reveal a
second order transition; although at $\beta=10.30$,
where we have calculations on two volumes, there is no
sign of the dip becoming deeper as the volume increases.
But we emphasise that our calculations were not
designed to locate a phase transition, and instead were
merely intended to identify the cross-over region.

Several reasons have been given in the past for a crossover 
or phase transition separating the strong and weak coupling 
regimes in SU($N$) gauge theories. One explanation focuses
on string roughening (see
\cite{Kogut,Creutz}
and references therein): in strong coupling the confining
string is rigid, as though the transverse fluctuations were
massive, while in weak coupling we recover the usual
continuum-like string with massless transverse fluctuations.
In the region of couplings where this mass vanishes the
string becomes rough and the lattice theory should 
exhibit a cross-over
\cite{Kogut,Creutz}.
A quite different approach
\cite{GW80}
focuses on the different eigenvalue 
distributions that the plaquette matrix possesses at strong
and weak coupling. From this one can conclude
\cite{GW80}
that there is a third-order phase transition separating
the strong and weak coupling regions at $N=\infty$ with,
presumably, some kind of cross-over at finite but large $N$. 
A third approach 
\cite{funadj,Creutz}
focuses on the phase structure in an extended coupling space,
driven by the condensation of various lattice topological
objects
\cite{BKL,Halliday}.
We will now comment on, and extend a little further,
this last idea.

We start by considering SU(2). Suppose one generalises
the lattice action in eqn(\ref{eqn_action}) to
\be
S = 
- \beta \sum_p \biggl (1 - {1\over N} ReTr \, U_p \biggr )
- \beta_A \sum_p\biggl (1 - {1\over{N^2-1}} ReTr_A \, U_p \biggr ),
\label{eqn_actionFA}
\ee
where  $Tr_A$ is the trace in the adjoint representation, and
the $Tr$ in the first term continues to be in the fundamental
representation. This lattice theory has a non-trivial phase 
structure in the $(\beta,\beta_A)$ space of couplings
\cite{funadj,Creutz}.
In particular there is a phase transition line which ends in a
critical point and this point approaches the Wilson axis,
$\beta_A = 0$, as we go from SU(2) to SU(3). Moreover if
one extrapolates this phase transition line one crosses
the Wilson axis in the weak-to-strong coupling crossover 
region. Since the mass gap vanishes at the critical point, 
its proximity to the Wilson axis provides an explanation for 
the dip in the mass of the scalar glueball in the region 
that characterises the transition from weak to strong coupling,
and also the fact that this dip deepens as we go from
SU(2) to SU(3). 

The SU(2) phase structure described above is believed to 
be driven by the following dynamics
\cite{BKL,Halliday}.
If we multiply a link matrix by an element of the centre
then this is invisible to the adjoint piece of the action.
Strings of plaquettes of flipped sign can be thought
of as $Z_2$ vortex lines. Such vortices may be closed
or may end on $Z_2$ monopoles. At small $\beta$ and small 
$\beta_A$ there will be a vacuum condensate of both the $Z_2$ 
vortices and the $Z_2$ monopoles. If we increase $\beta_A$
at $\beta \sim 0$ the vortices remain condensed since
the adjoint action does not feel their presence.
However the monopoles will cease to condense at some
critical value of the coupling for the same entropy/action
reason that U(1) monopoles cease to be condensed in
the D=3+1 U(1) theory beyond a certain critical coupling.
Thus there is a phase transition line extending to finite
$\beta$ from some point along the adjoint axis.
Now if we go to large $\beta_A$ the plaquettes are
all close to $\pm 1$ and we have something close to
a $Z_2$ spin system. If we then increase $\beta$ 
this is equivalent
to increasing an external field that breaks the  $Z_2$
symmetry and at some point there is a phase transition
to a phase where the (ultraviolet)  $Z_2$ vortices
are suppressed. So there is a phase transition line
descending into the $(\beta,\beta_A)$ plane. At some
point this coalesces with the other phase transition
line and continues downwards till it ends at the
critical point. That it must end somewhere follows from 
the fact that at large negative $\beta_A$ the plaquette
traces are driven close to zero, and such a vacuum 
will support  neither $Z_2$ vortices nor monopoles.

We note that although this kind of explanation
sounds very particular to the simple plaquette action,
one should be able to generalise it to any loops;
the crucial thing is that we consider them in both
the fundamental and adjoint representations. One
might wonder whether other representations might be
relevant. It is clear however that the relevant
topological objects are determined by the centre of
the group and a loop in any SU(2) representation 
responds either like the fundamental or adjoint
representation to centre elements of the group.
Thus the analysis possesses a qualitative universality.

In SU(3) the picture is essentially the same,
but if we go to SU(4) things become different.
Now we can introduce an extra $k=2$ coupling, $\beta_2$, 
corresponding to double flux plaquettes. Such plaquettes
are insensitive to a factor $e^{i\pi}=-1$, just like
the adjoint coupling. However the adjoint action is
also insensitive to $e^{i\pi/2}$ fluxes. By the above
kind of argument we expect a similar phase structure
in the SU(4) $(\beta,\beta_{k=2})$ plane to the one we had 
in the SU(2) $(\beta,\beta_A)$ plane. Indeed we should
now consider the $(\beta,\beta_{k=2},\beta_A)$ manifold
of couplings, and there will be surfaces of phase 
transitions and lines of critical points which may
approach the Wilson axis at more than one location.
For higher SU($N$) we have a 
$(\beta,\beta_2,\beta_3,\ldots,\beta_A)$ 
space of couplings where the different pieces
of the action will be insensitive to different
elements of the centre. That is to say, to 
monopoles and vortices of various multiples of the
basic $Z_N$ flux. There will be a complex phase structure
in this space, but since it involves the condensation
of ultraviolet objects, one can in principle estimate 
the location of these transitions. We do not attempt to do so 
here. However we point out that these phase transitions 
would naturally be related to the eigenvalue spectrum
of the typical SU($N$) plaquette. It is not unreasonable
to conjecture that the smooth $N\to\infty$ limit of this
complex phase structure might be precisely the third-order 
phase transition discussed in
\cite{GW80}.

Finally we remark that there is no sign of a cross-over,
marked by a dip in the mass gap, when we consider
SU($N$) gauge theories in D=2+1. Since the vacuum
possesses all the topological objects described above,
this might look like a counterexample to the argument.
In fact this is not so. The key observation is
that in the D=2+1 U(1) theory there is no phase transition
between strong and weak coupling: in contrast to the
D=3+1 case. This is because in D=2+1 the monopoles
are instantons, while in D=3+1 they are objects with 
world lines. Thus the action/entropy balance is
quite different and leads to a quite different phase
structure. This applies equally well to $Z_N$ monopoles.

\section{The spectrum and string tension}
\label{sec_spectrum}

In Sections \ref{subsec_string} and \ref{subsec_masses} we 
describe our calculations of the string tension and of the
mass spectrum respectively. We then discuss,
in Section \ref{subsec_discuss}, what are the lessons
we learn as to how to go about doing a much better calculation.

\subsection{the string tension}
\label{subsec_string}

To calculate the (fundamental) string tension $\sigma$
we construct string operators that close upon themselves
through a spatial boundary. The correlation function of 
two such strings separated by a (Euclidean) time interval  
$t$ will, for large enough $t$, decrease exponentially 
$\propto \exp(-m_l t)$ where $m_l$ is the mass of the
lightest periodic flux loop. On the lattice $t=an_t$
where $n_t$ labels the temporal time-slices, so what
we obtain is the value of $am_l$, the mass in lattice units.
If the length of the loop is $aL$ and if the theory is 
linearly confining then $a^2\sigma = \lim_{L\to\infty} am_l/L$. 
Moreover the first correction to this linear dependence is 
universal
\cite{Luscherstring}
if the infrared properties of the confining flux tube are
described by an effective string theory. If we further assume 
that the universality class is that of a simple (Nambu-Gotto)
bosonic string (there is some evidence from previous SU(2) 
and SU(3) lattice calculations that points to this) and if we 
use the string correction appropriate to our closed periodic string
\cite{Polystring}
then we find
\be
a m_l = a^2 \sigma L - \frac{\pi}{3 L}
\label{eqn_poly}
\ee
once the flux loop is large enough. We shall use this expression
to extract $a^2\sigma$ from our lattice calculations of the
flux loop mass $am_l$.

\begin	{figure}[p]
\begin	{center}
\epsfig{figure=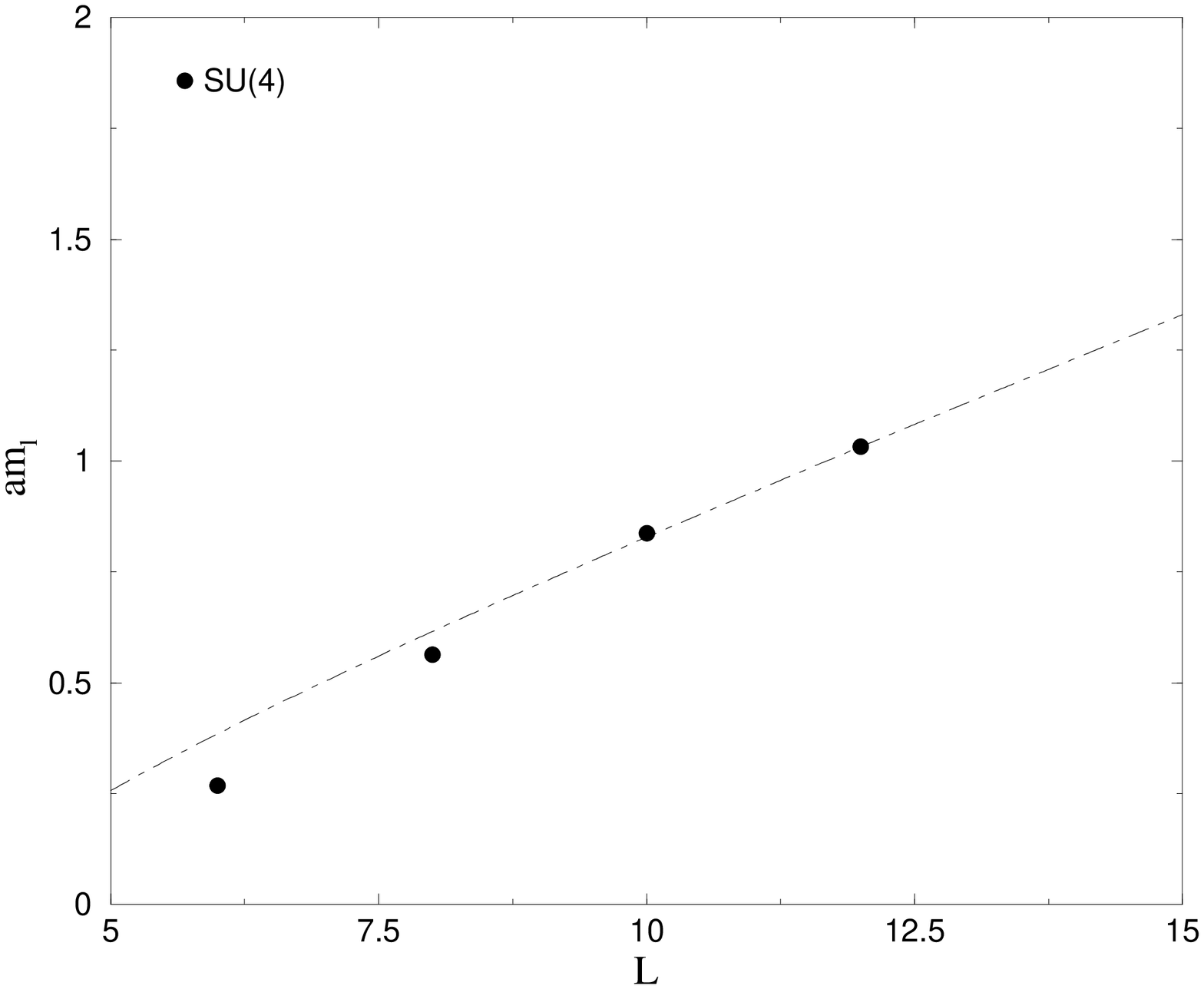, angle=270, width=15cm} 
\end	{center}
\vskip 0.15in
\caption{The mass of the lightest periodic flux loop as a function
of its length, at $\beta=10.70$ in SU(4). Shown is a linear fit 
with a string correction as in eqn(\ref{eqn_poly}).}
\label{fig_polymass4}
\end 	{figure}

\begin	{figure}[p]
\begin	{center}
\leavevmode
\epsfig{figure=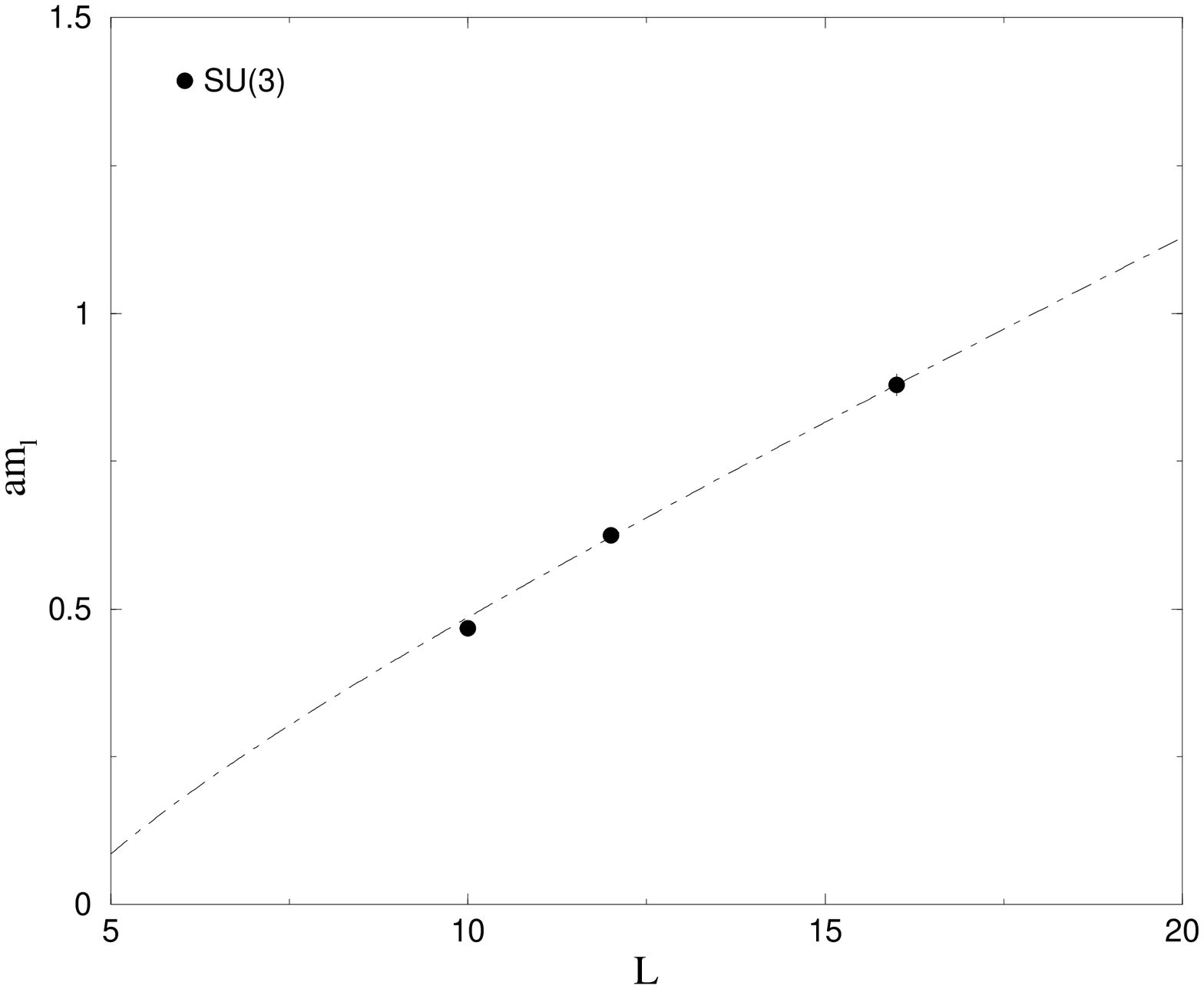, angle=270, width=15cm} 
\end	{center}
\vskip 0.15in
\caption{The mass of the lightest periodic flux loop as a function
of its length, at $\beta=5.93$ in SU(3). Shown is a linear fit 
with a string correction as in eqn(\ref{eqn_poly}).}
\label{fig_polymass3}
\end 	{figure}

As we have just remarked, in using eqn(\ref{eqn_poly}) we are 
assuming that we have linear confinement in our $SU(N)$ 
gauge theories, that the effective string theory describing the
infrared properties of the confining flux tube is in the 
Nambu-Gotto universality class, and that our string is in fact long
enough for higher order corrections to be negligible. There is,
of course, a great deal of numerical evidence, scattered through
the literature, for linear confinement in the case of SU(2) and SU(3) 
as well as some (much weaker) evidence that the leading correction is 
as in eqn(\ref{eqn_poly}). To address these issues for $SU(N>3)$ 
we show in Fig.\ref{fig_polymass4} how
the value of $am_l$ varies with $L$ in SU(4) at a particular
value of $a$. In Fig.\ref{fig_polymass3} we perform a similar
exercise for the case of SU(3). We observe that for SU(4), just as 
for SU(3), the mass $am_l$ increases roughly linearly with $L$:
evidence for linear confinement. We also note that in both cases
the deviation from linearity is consistent with eqn(\ref{eqn_poly})
for the largest two values of $L$ and that in both cases this occurs 
for distances $aL \geq 3/\surd\sigma$. If we equate the masses
corresponding to these longest two loops to $a m_l = bL -c/L$
then we find $c=0.94\pm 0.41$ and  $c=0.76\pm 0.43$ for
the SU(3) and SU(4) cases respectively. These values
are consistent with the value $c= \pi/3 \simeq 1.05$ assumed
in eqn(\ref{eqn_poly}). We emphasise that, with only
two values of $L$ fitting this correction, we cannot claim
to have evidence for both the $1/L$ functional form and for its
coefficient. Rather we can say that if we assume the functional
form then we have some evidence that the bosonic string
coefficient, as in eqn(\ref{eqn_poly}), is the correct one.
This is thus a minimal finite size study (which we are in
the process of improving upon
\cite{blmtstring2})
but it does indicate 
that it should be safe, within our statistical accuracy, to use 
eqn(\ref{eqn_poly}) to extract $a^2 \sigma$ as long as we do so 
for $aL \geq 3/\surd\sigma$.

We remark that although our determination of the string
correction is not very precise, using periodic loops rather
than Wilson loops is by far the better way to approach this 
question. One reason is that the string correction has a 
larger coefficient in the former case; $c=\pi/3$ as compared
to $c=\pi/12$ for Wilson loops. More importantly, with Wilson 
loops the string is attached to static sources which provide 
a Coulomb interaction. This has the same functional form as 
the string correction and dominates the potential at shorter 
distances. Thus in fits to the potential of the form
$V(r) = v + \sigma r + c/r$ the value of $c$  will be
dominated by the short-distance Coulomb interaction rather
than the long-distance string correction, unless one confines
the fitting range to distances greater than, say, 1 fm. 
(Indeed, as we have seen, the periodic flux loop has
to have a length  $r \geq 3/\surd\sigma \simeq 1.3fm$
if the leading string correction is to dominate.) 
This is a tough criterion for Wilson loop calculations
and is rarely met.

The loop masses and string tensions obtained in our
SU(2), SU(3), SU(4) and SU(5) lattice calculations
are listed in Tables \ref{table_datsu2}, \ref{table_datsu3}, 
\ref{table_datsu4} and \ref{table_datsu5} respectively.
In the various continuum extrapolations that
we shall perform we shall only use string tensions calculated
on lattices that satisfy $aL \geq 3/\surd\sigma$ and thus ones
to which we believe the application of eqn(\ref{eqn_poly})
is justified.

\subsection{the mass spectrum}
\label{subsec_masses}

In addition to the string tension, we calculate the masses
of the lightest scalar and tensor glueballs, as well as
the mass of the first scalar excitation. These are obtained
from correlations of operators that are obtained by 
multiplying link matrices around closed contractible loops.
These masses are listed in Tables \ref{table_datsu2}, 
\ref{table_datsu3}, \ref{table_datsu4} and \ref{table_datsu5}.

The first question to ask is how large the spatial volume has 
to be for glueball finite volume corrections to be negligible
(at our level of statistical accuracy). To answer this question
we have performed mass calculations on a range of
lattice volumes at $\beta=10.7$ in the case of SU(4) and
at $\beta=5.93$ in the case of SU(3). The masses are listed in
Tables \ref{table_datsu4} and \ref{table_datsu3} respectively.
We observe that the lightest $J^{PC}=0^{++}$ and $2^{++}$
glueballs appear to have reached (within errors) their infinite 
volume limit once $aL\surd\sigma \geq 3$. For the excited
$0^{++\star}$ scalar the situation is less clear: it appears
to be the case for SU(4) but apparently not for SU(3). We
shall return to the probable cause of this in 
Section \ref{subsec_discuss} but for now we shall simply assume
that it is safe to calculate glueball masses and the string
tension on lattice volumes that satisfy the constraint
$L \geq 3/a\surd\sigma$.

We are interested in determining the $N$ dependence
of physical quantities in the continuum limit. We first note
that if we form a dimensionless ratio of physical quantities
then the lattice scale drops out, e.g. 
$am_G/a\surd\sigma \equiv m_G/\surd\sigma$. This ratio
will of course still possess lattice corrections. However the
functional form of such corrections is known. In particular
for our simple plaquette action we expect the leading
correction to be $O(a^2)$. Thus for small enough $a$
we expect the continuum limit to be approached as
\be
{{m_G(a)} \over {\surd\sigma(a)}} =
{{m_G(0)} \over {\surd\sigma(0)}} + c a^2\sigma.
\label{eqn_cont}
\ee
We could use any mass $\mu$ in the correction term,
since $a^2\sigma$ and $a^2\mu^2$ differ at $O(a^4)$.
We choose to use $\sigma$ both for the correction
term and in our dimensionless mass ratios, because
it is the quantity we calculate most accurately
and reliably. (In reality $c$ in eqn(\ref{eqn_cont})
is a power series in $g^2(a)$. However, since the
logarithmic variation of the coupling is very small
over our range of $a$, we follow usual practice and
treat it as a constant.)

Our procedure is therefore to fit our lattice ratios 
with eqn(\ref{eqn_cont}). We begin by including all
our calculated values. If the best fit is poor, we drop 
the coarsest value of $a$ and re-attempt the fit -- and 
so on. Following this procedure we obtain the continuum 
mass ratios listed in Table \ref{table_continuumG}. 
We remark that the SU(2) and SU(3) values agree with those 
obtained previously, as reviewed for example in
\cite{MTrev98},
and that our SU(2) mass ratios are much more accurate than
earlier work, as expected. 

\begin	{figure}[p]
\begin	{center}
\epsfig{figure=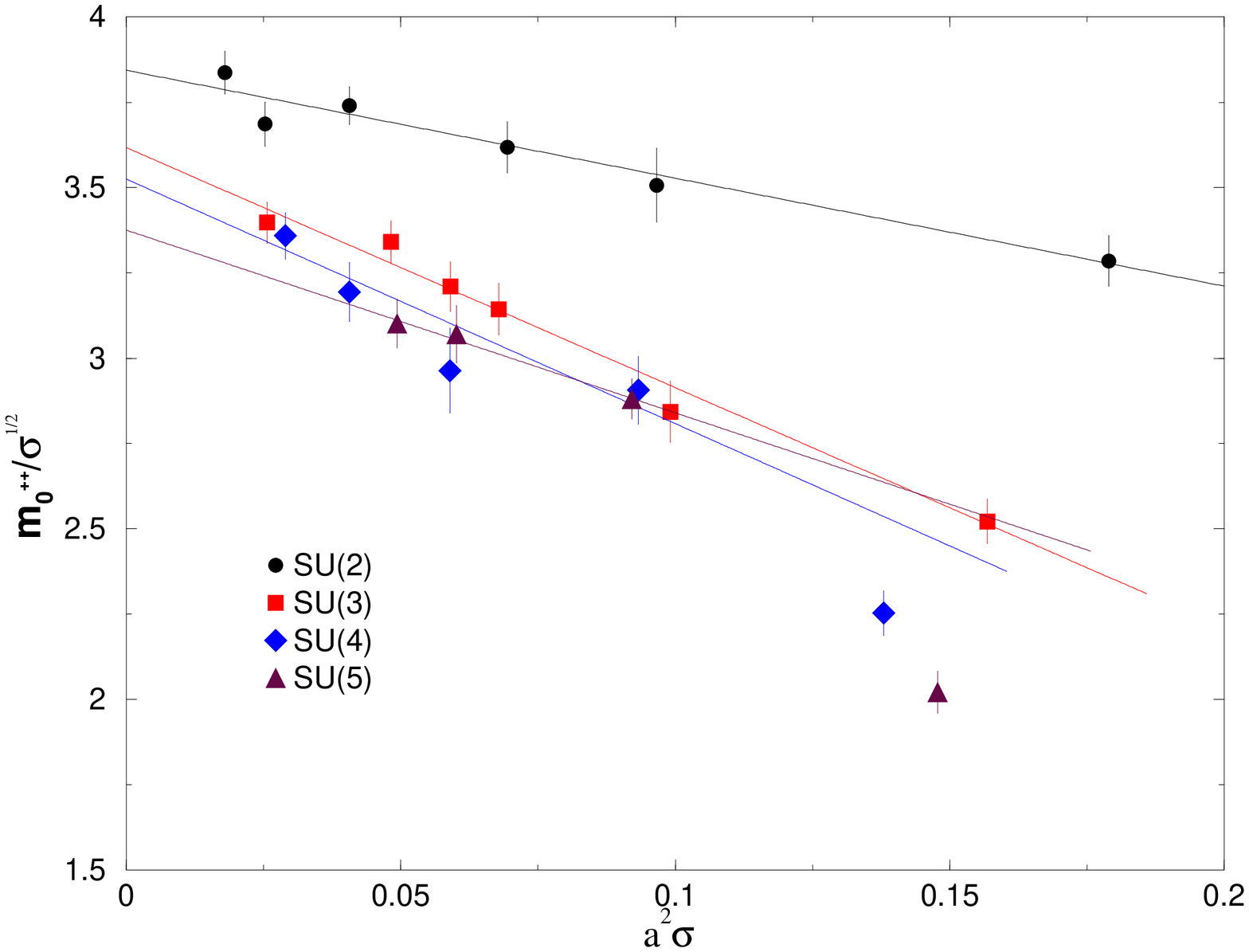, angle=270, width=15cm} 
\end	{center}
\vskip 0.15in
\caption{The mass of the lightest scalar glueball, $m_{0^{++}}$, 
expressed in units of the string tension, $\sigma$, is plotted 
against the latter in lattice units. The continuum extrapolation, 
using a leading lattice correction, is shown.}
\label{fig_scalar}
\end 	{figure}

\begin	{figure}[p]
\begin	{center}
\epsfig{figure=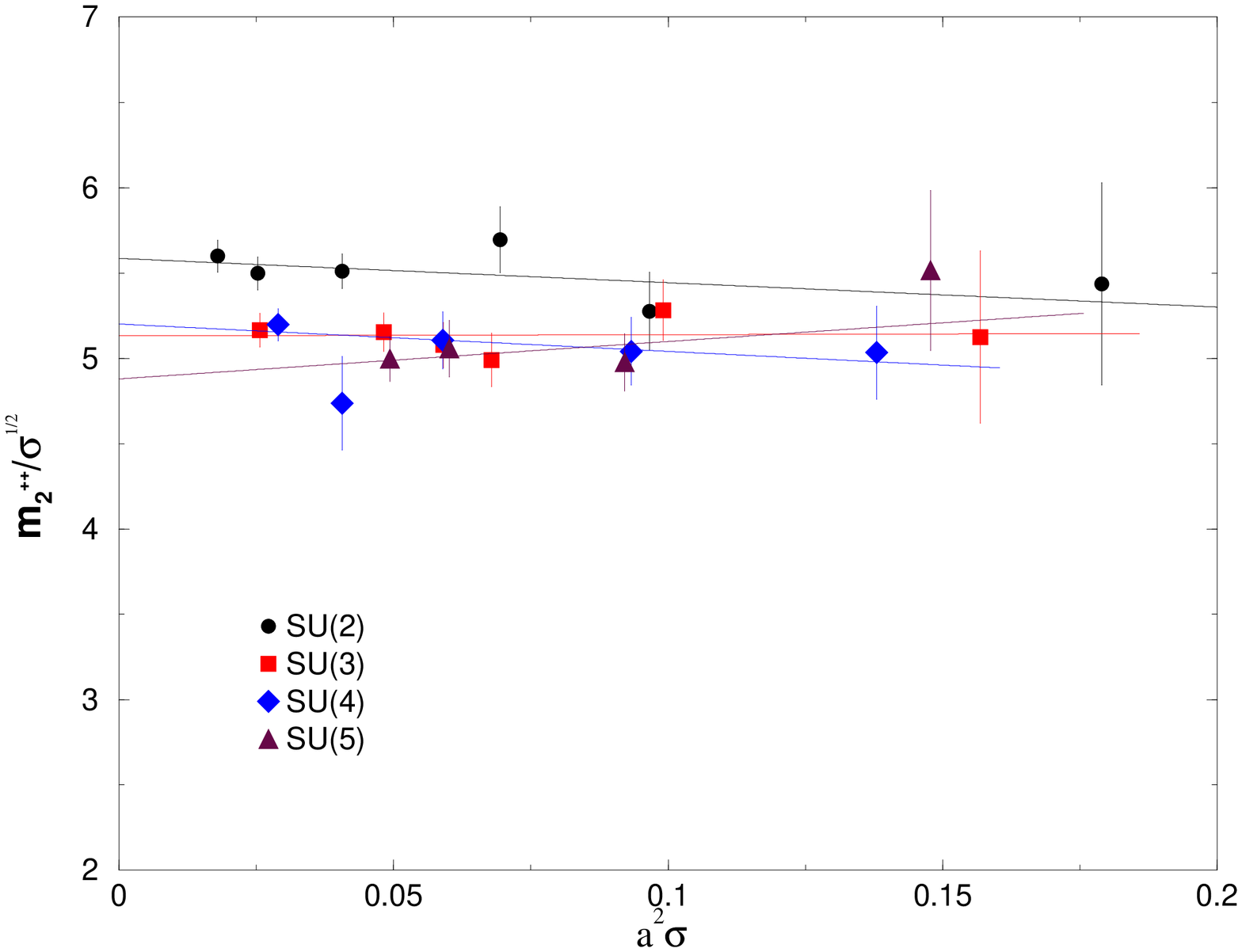, angle=270, width=15cm} 
\end	{center}
\vskip 0.15in
\caption{The mass of the lightest tensor glueball, $m_{2^{++}}$, 
expressed in units of the string tension, $\sigma$, is plotted 
against the latter in lattice units. The continuum extrapolation, 
using a leading lattice correction, is shown.}
\label{fig_tensor}
\end 	{figure}

The best fits involved in the extrapolations of the
lightest $0^{++}$ and  $2^{++}$ glueballs are almost
always very good, and in any case never poor. They
are shown, together with the lattice ratios, in
Fig.\ref{fig_scalar} and Fig.\ref{fig_tensor}. 
We note the increasing influence as $N$ increases of 
the strong-to-weak coupling crossover on the mass of 
the $0^{++}$ at the coarsest lattice spacing. For
SU(5) the $0^{++\star}$ fit is also good, and for SU(2)
it is at least acceptable. However for SU(3) and SU(4) 
the best fits are very poor. In these cases we have
quoted generous errors in an attempt not to be
misleading. We believe we understand the origin of
these problems with the $0^{++\star}$, and this
is discussed in Section \ref{subsec_discuss}.

\begin	{figure}[p]
\begin	{center}
\epsfig{figure=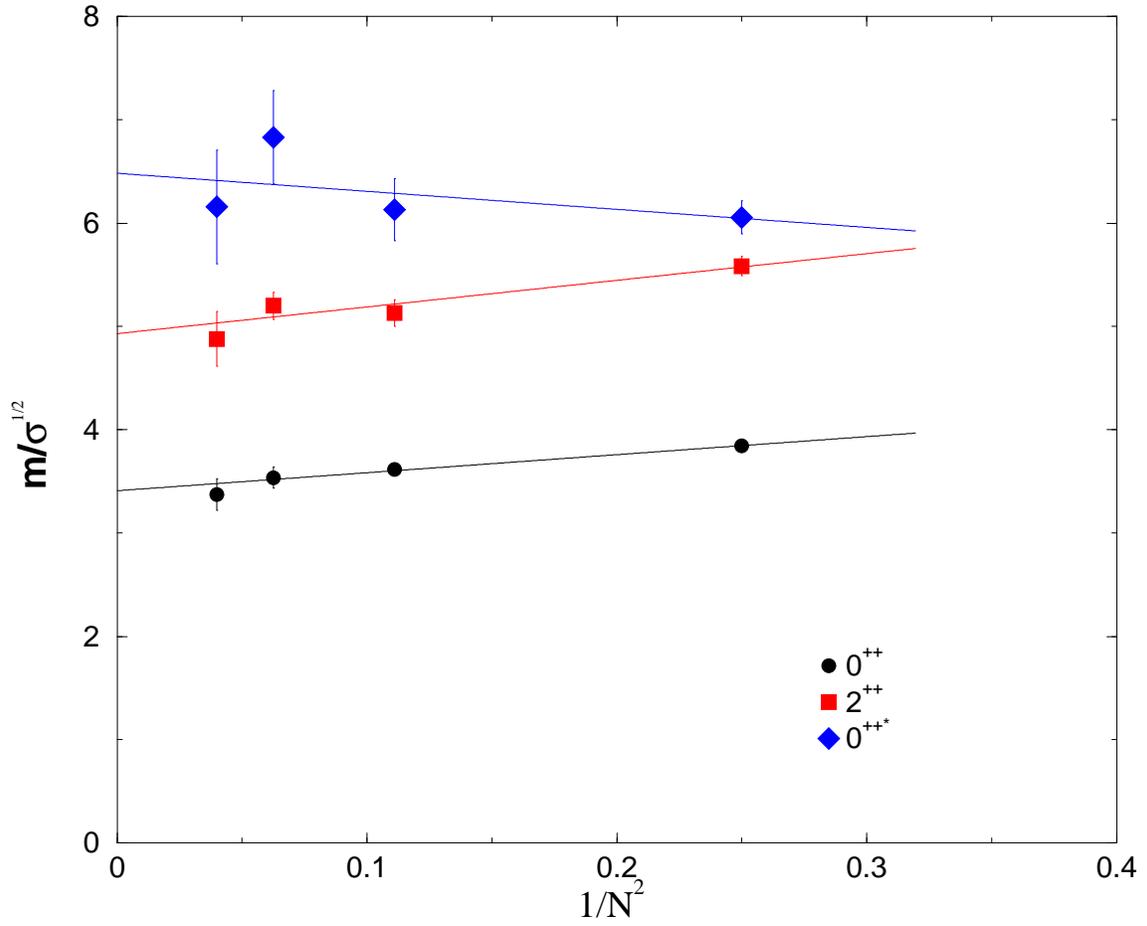, angle=270, width=15cm} 
\end	{center}
\vskip 0.15in
\caption{Continuum scalar, tensor and excited scalar
masses expressed in units of the string tension and plotted
against $1/N^2$. Linear extrapolations to $N=\infty$
are shown in each case.} 
\label{fig_glueN}
\end 	{figure}

We are now in a position to discuss the $N$
dependence of continuum SU($N$) gauge theories.
We expect that once $N$ is large enough, mass
ratios will have a smooth limit with a leading
correction that is $O(1/N^2)$, i.e. 
\be
\left.  {{m_G} \over {\surd\sigma}} \right |_{N} =
\left.  {{m_G} \over {\surd\sigma}} \right |_{\infty} 
+ {c \over N^2}.
\label{eqn_largeN}
\ee
In Fig.\ref{fig_glueN} we plot the continuum mass ratios 
against $1/N^2$. According to eqn.(\ref{eqn_largeN})
once $N$ is sufficiently close to $N = \infty$
the mass ratios should fall on a straight line. 
Performing linear fits we observe that in fact this is
the case all the way down to $N=2$. For the $0^{++\star}$
this is perhaps not so significant because the errors are 
so large; however in the case of the $2^{++}$ and
of the $0^{++}$ the errors are small and the result is
striking. 

Our best fits for $N \geq 2$ are:
\be
{{m_{0^{++}}} \over {\surd\sigma}}  
=
3.37(15) + { {1.93(85)} \over N^2}
\label{eqn_scalarN}
\ee
\be
{{m_{2^{++}}} \over {\surd\sigma}}  
=
4.93(30) + { {2.6(1.9)} \over N^2}
\label{eqn_tensorN}
\ee
and
\be
{{m_{0^{++\star}}} \over {\surd\sigma}}  
=
6.43(50) - { {1.5(2.6)} \over N^2}
\label{eqn_excscalarN}
\ee
These relations give us the corresponding mass ratios
not only for $N=\infty$ but for all values of $N$.
The fact that we can fit these mass ratios with just the 
leading $1/N^2$ corrections all the way down to $N=2$
and the fact that the coefficients are quite small
can be summarised by saying that all SU($N$) gauge
theories are close to  SU($\infty$), at least as far
as the low-lying mass spectrum is concerned.

\subsection{discussion}
\label{subsec_discuss}

Our above results provide a clear motivation for a
much more detailed calculation. From what we have 
learned here, how should such a calculation proceed? 

We have seen that there is a large gap to the first
excited scalar state. The lesson is that if we
want a detailed mass spectrum, many of the states will
be quite heavy in lattice units. In this situation
the best strategy is to use an anisotropic lattice 
with a fine temporal lattice spacing that allows a 
much finer resolution of the rapidly decreasing
correlation functions that correspond to heavy masses.
This is an idea and technique with a long history
\cite{MTanisotropic}
which has been used to impressive effect in recent
calculations of the SU(3) mass spectrum
\cite{MPanisotropic}
(and also in calculations of the static potential
\cite{BaliCS}).

The first excited scalar has a mass 
$m_{0^{++\star}} \simeq 2 m_{0^{++}}$. This raises
the question whether it might not in fact be a
two glueball scattering state with zero relative
momentum. Within our present calculation we cannot 
answer this question with any certainty. Any
future calculation should include in the variational
basis of operators explicit multi-glueball operators
for all the scattering states, of various relative
momenta, that are expected to lie in the mass
range of interest. In this way we can hope to identify
and separate bound states or resonances from the
less interesting scattering states. 

In addition to any bound states, resonances and scattering
states, there are also extra states which decouple in
the infinite volume limit but which are not so heavy that
they can be ignored for the kind of volumes we are likely
to use. These `torelon' states (see
\cite{mt98}
and references therein), which are composed 
of a periodic flux loop and its conjugate, have
a mass that is about twice the mass of the single 
flux loop: $m_T \simeq 2 m_l$. This mass increases
roughly linearly with
the volume of course, but for much of the present
calculations this happens to be close to $m_{0^{++\star}}$.
Indeed the strong finite-size effects we observe 
$m_{0^{++\star}}$ to possess in our SU(3) study, 
suggests an occasional misidentification with such
a torelon state. 
In the case of SU(3) and SU(4) the volumes are less
constant than in the case of SU(2) which suggests
this as the reason for the poor continuum extrapolation.
Again the solution to the problem is to include
explicit torelon operators in the variational basis
so as to make the explicit identification of these
states possible.

The above discussion should however not be taken to
imply that we have no faith in our estimate of
the excited scalar mass. The reason is that
our glueball operators involve the trace of a single loop,
and we expect the overlap of both the torelon and two-glueball 
states on such operators to decrease rapidly with increasing 
$N$. Thus we are inclined to believe that what we observe
in SU(5) is a genuine excited glueball state, so that
our large-$N$ mass estimate should be quite reliable.

Our second observation, based on our tabulated masses,
is that the errors on the masses appear to increase
as we go from SU(2) to larger $N$. This could be a
problem that is intrinsic to such a calculation. One
possible reason is that, as we shall see below, 
the sampling of different vacuum topological sectors
becomes rapidly less ergodic as $N$ increases. Another
reason is that as $N$ grows the path integral is
increasingly dominated by the single `Master Field'
\cite{largeN}.
Our mass calculations are derived from the correlations 
in distant fluctuations about this Master Field and perhaps
this method becomes inefficient at large $N$. (In which case
it is an interesting question to ask how one might
calculate the masses.) Because of its practical importance
we have investigated this question in detail. What we find
is that the actual correlation functions do not display
statistical errors that grow with $N$. This is good news: it
means that the problem is not an intrinsic one. What happens
is that our best operators have a poorer overlap onto
the lightest states as $N$ increases. This is more pronounced
for the flux loop than for the glueballs, and becomes more
pronounced at smaller $a$ in the latter case. It may be useful
to illustrate this with some explicit values. Consider the
overlap that appears in eqn(\ref{eqn_corrln}),
$c^2_n \equiv |\langle \Omega | \phi | n \rangle |^2$,
where we normalise $\sum_n c^2_n =1$. For the best 
$0^{++}$ glueball operator the value of $c^2_n$ varies,
for our coarsest common value of $a$, from $\sim 0.95$ in
SU(2) to $\sim 0.90$ in SU(5). For the finest (common)
values of $a$ it varies from $\sim 0.92$ to $\sim 0.85$.
For the periodic flux loop the corresponding variation
of $c^2_n$ is from $\sim 0.90$ to $\sim 0.75$ for both
coarse and fine values of $a$. This behaviour suggests
that the problem is with our simple blocking algorithm.
A better strategy would be to intermingle smearing and
blocking steps. In any case the lesson here is that a much  
better calculation will require the construction of operators 
with better overlaps.

Another important issue concerns quantum numbers.
Typically our operators fall into representations
of the cubic group. What we have called $0^{++}$ is
actually the cubic $A^{++}_1$ representation which 
also includes pieces of other continuum representations,
such as the $4^{++}$. For the lightest states one
can identify criteria 
\cite{mt98}
which reassure us that our continuum spin labelling
is correct. However the ambiguity becomes larger
for excited states. For example, one can ask whether
our claimed $0^{++\star}$ might not in fact be the
lightest $4^{++}$. Since the higher-spin continuum
representations are typically spread between
several representations of the cubic group, one
approach to resolving these ambiguities is to 
search for corresponding degeneracies
between states in different cubic representations.
However it is unlikely that one will, in practice, obtain 
these heavier masses with sufficient accuracy to establish
degeneracy unambiguously. (Which, in any case, only becomes
exact for small $a$.) An alternative (or additional) strategy
is to construct operators with approximate continuum
rotational properties
\cite{mtrj}.
One can then use the size of the overlap of the excited states
onto such operators to determine what their likely quantum 
numbers are.

\section{'t Hooft coupling}
\label{sec_coupling}

The calculations in the previous Section indicate that SU($N$) 
gauge theories have a smooth limit as $N\to \infty$, that the 
theory remains confining, and that the approach is `precocious' 
in the sense that even SU(2) appears to be within a 
modest $O(1/N^2)$ correction of SU($\infty$). 

The analysis of diagrams 
\cite{largeN}
suggests that such a smooth limit should
be achieved by keeping the 't Hooft coupling, 
$\lambda \equiv g^2 N$, constant as $N \to \infty$. We shall
now see to what extent this expectation can be addressed by
our calculations. 

Since the coupling runs, the 't Hooft coupling is not a constant
but will depend on the length scale $l$ on which it is defined,
i.e. $\lambda(l) \equiv g^2(l) N$. Thus we expect that
if we fix the value of $l$ in units 
of some quantity that partakes of the smooth large-$N$ limit,
such as the string tension,
then $\lambda(l)$ should have a smooth non-trivial limit
as $N \to \infty$. Now, we can
define couplings in various ways, and one way is to use
$\beta = 2N/g^2$. Here $g^2$ is the lattice bare coupling and
provides a definition of a coupling on the length scale $a$. 
Thus we define 
\be
\lambda(a) =  g^2(a) N = 2N^2/\beta
\label{eqn_lambda}
\ee
However it is well known that in our range of $a$ this
coupling is heavily influenced by lattice artifacts peculiar
to the plaquette action, and that a much better coupling
can be obtained from it by the definition
\be
\lambda_I(a) =  g_I^2(a) N 
= 2N^2/\beta_I 
= 2N^2/(\beta \times {1\over N}\langle ReTr \, U_p \rangle )
\label{eqn_lambdaI}
\ee
which may be regarded as a mean-field improved or tadpole
improved version of $\beta$ and $\lambda(a)$
\cite{gimp}.

An aside. Ideally we would have wished to define a coupling on 
some scale $l \gg a$ so that we could determine its behaviour 
at fixed $l$ as $a\to 0$ and then compare how this
continuum coupling varies with $N$. (For example 
one might use the coupling definition in
\cite{alpha}.)
Using $\beta$, as we do, mixes in lattice corrections in an uncontrolled 
fashion. Transforming this to $\beta_I$ is known to remove a large
part of the lattice corrections at small $a$. Nonetheless, the fact 
that these lattice corrections vary with the scale $l$, because 
of course $l=a$, will limit our ability to probe the interesting
question of what happens to the coupling on larger
distance scales.
 
\begin	{figure}[p]
\begin	{center}
\epsfig{figure=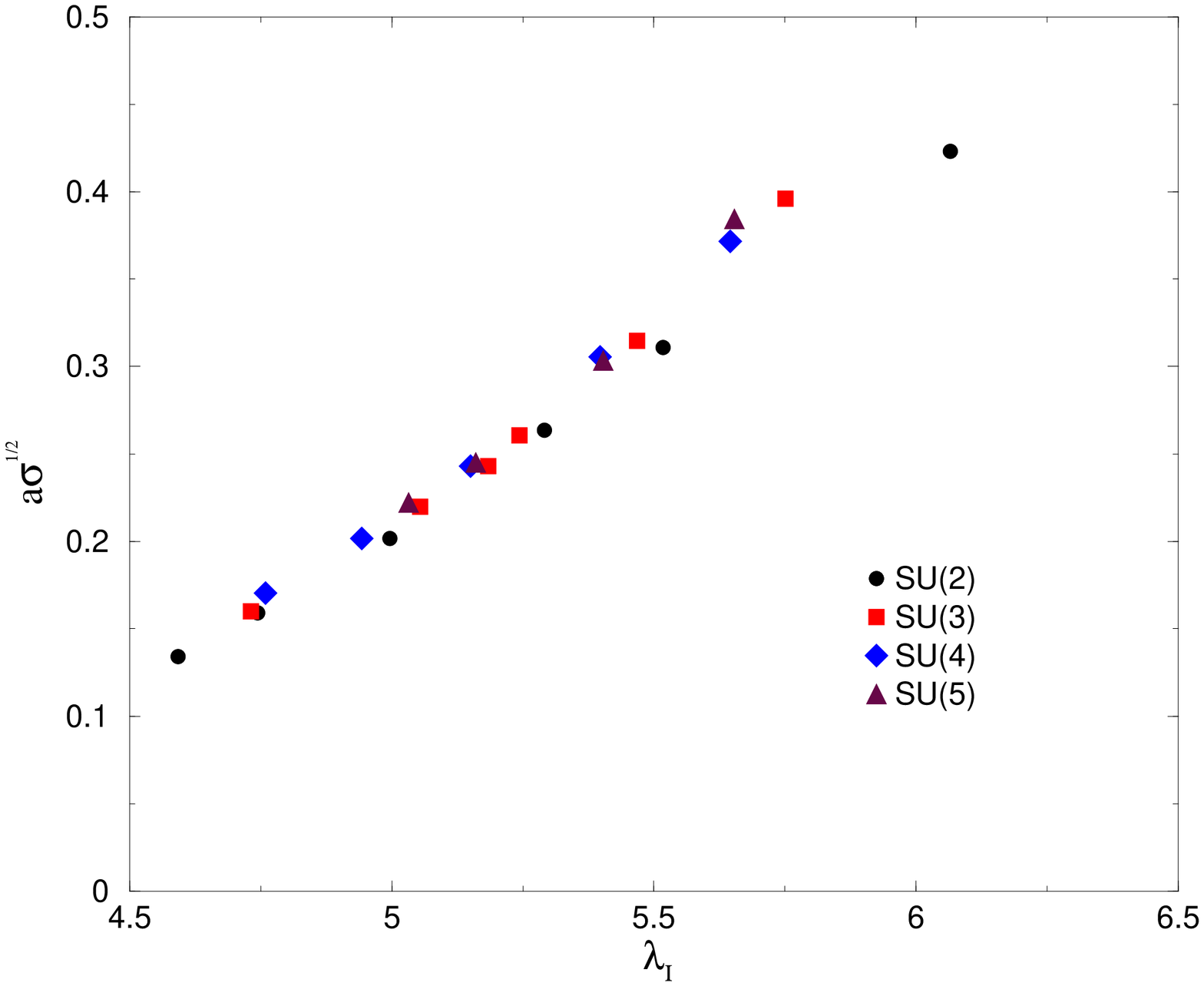, angle=270, width=15cm} 
\end	{center}
\vskip 0.15in
\caption{The square root of the string tension in lattice units,
$a\surd\sigma$, plotted against the 't Hooft coupling, 
$\lambda_I \equiv g^2_I N$.}
\label{fig_betaI}
\end 	{figure}

To test whether we get a smooth large-$N$ limit by keeping 
$\lambda_I(l=a)$ fixed, we need to choose a physical length scale 
in terms of which we keep $l$ fixed.  We shall 
choose $1/\surd\sigma$ since we have already seen that the ratio 
of $\surd\sigma$ to other masses has a smooth large-$N$ limit.
Thus our first expectation is that if we plot  $a\surd\sigma$ 
against $\lambda_I(a)$ then, for large enough $N$, the calculated 
values should fall on a universal curve. In Fig.\ref{fig_betaI}
we plot all our calculated values of $a\surd\sigma$,  against
the value of  $\lambda_I(a)$, for $N=2,3,4$ and $5$. 
We observe that, within corrections which are small except at
the largest value of $a$, we do indeed see a universal curve.

\begin	{figure}[p]
\begin	{center}
\epsfig{figure=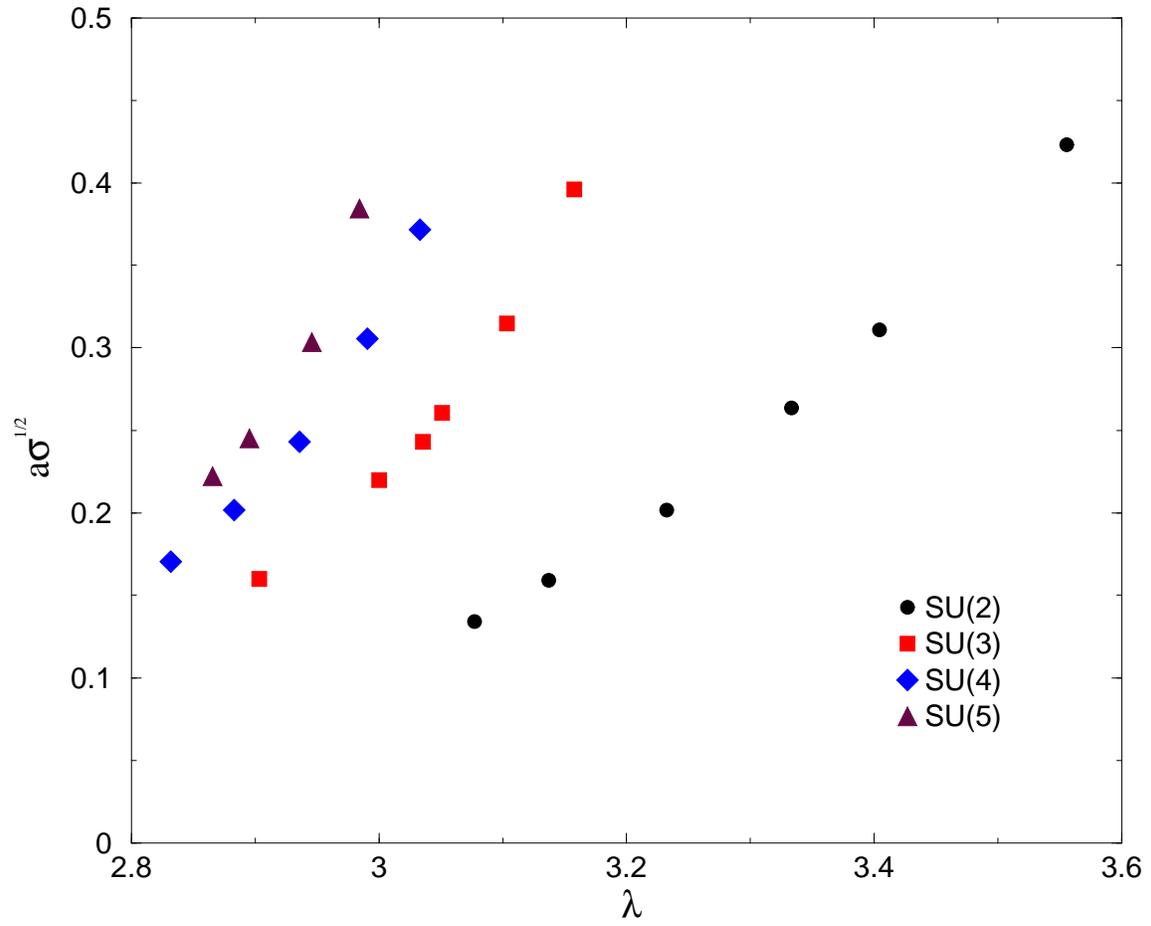, angle=270, width=15cm} 
\end	{center}
\vskip 0.15in
\caption{The square root of the string tension in lattice units,
$a\surd\sigma$, plotted against the unimproved 't Hooft coupling, 
$\lambda \equiv g^2 N$.}
\label{fig_beta}
\end 	{figure}

It is instructive to ask what would have happened if we had used
eqn(\ref{eqn_lambda}) rather than eqn(\ref{eqn_lambdaI}) to
define our running coupling. The answer is displayed
in Fig.\ref{fig_beta}. Here we see how very large lattice
corrections can easily hide the approach to large-$N$.

\begin	{figure}[p]
\begin	{center}
\epsfig{figure=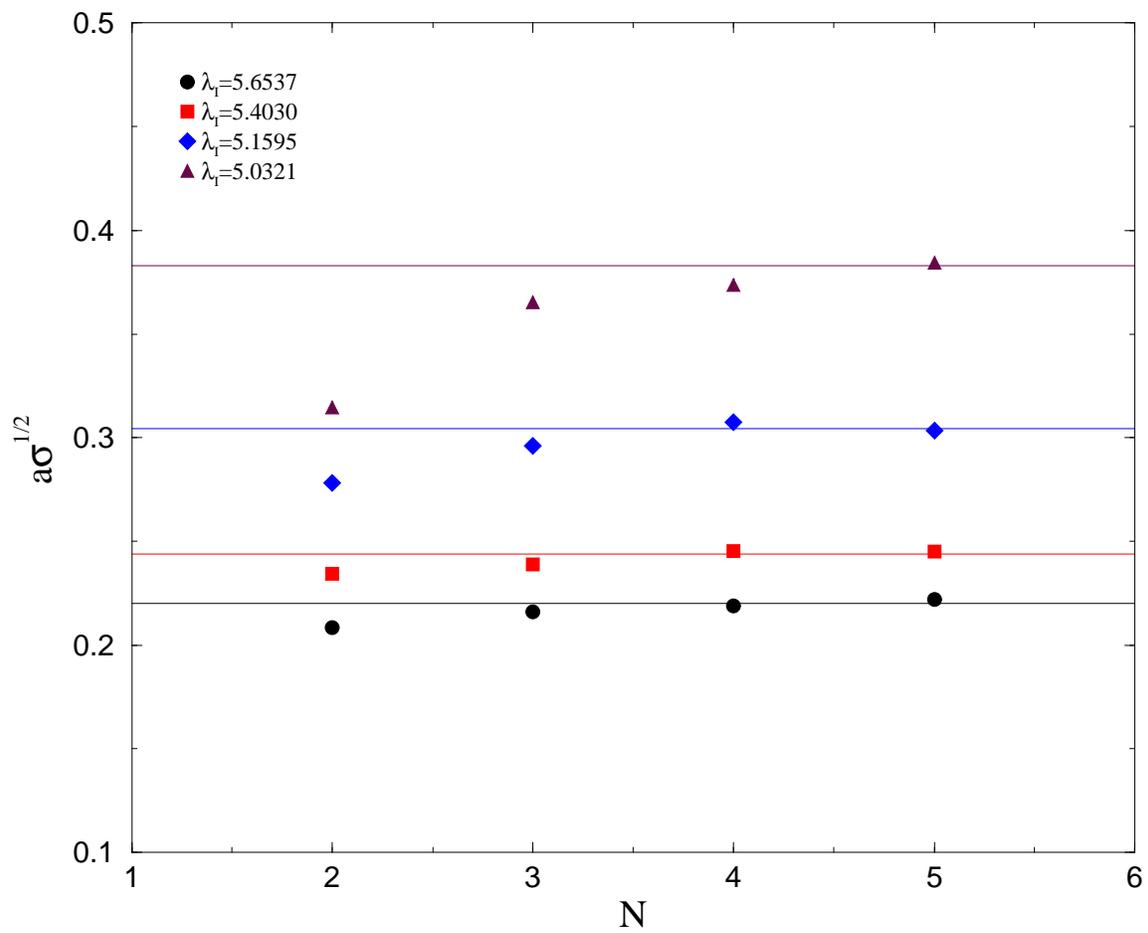, angle=270, width=15cm} 
\end	{center}
\vskip 0.15in
\caption{We show how the string tension varies with $N$ at
four different values of the 't Hooft coupling.}
\label{fig_coupling}
\end 	{figure}

While Fig.\ref{fig_betaI} makes the qualitative point, it is hard 
to make out the details of the approach to the $N=\infty$ limit.
To display this we start with each of our four SU(5) values 
of $\beta$ and calculate the value of $a\surd\sigma$ obtained
at the corresponding value of $\lambda_I$ for $N =2,3,4,5$. 
(To obtain these values for $N \not = 5$ requires some 
interpolation.) In Fig.\ref{fig_coupling} we show how the string 
tension varies with $N$ at these four fixed values of the 't Hooft 
coupling. We see that at all values of $a$ except the largest
(where we are close to the strong-to-weak coupling crossover) 
the string tension at fixed 't Hooft coupling becomes independent 
of $N$, within our errors, for $N \geq 4$.

Once the coupling $g_I^2(a)$ is small enough to be accurately
given by its 2-loop perturbative formula, we can replace
the above analysis by simply extracting the
scale $\Lambda_I$ that appears in the 2-loop formula
for the running coupling, and seeing if it also has a
smooth large-$N$ limit. We can use our calculations to illustrate 
how such an analysis proceeds, even though we do not really 
expect our couplings to be small enough for it to be very reliable.
First we invert the 2-loop formula so
as to obtain $a$ in terms of $g^2(a)$, and then we multiply
both sides by $\surd\sigma$ to obtain
\be
a\surd\sigma ={{\surd\sigma}\over{\Lambda_I}}
\biggl({{24\pi^2}\over{11}}{{\beta_I}\over{N^2}}\biggr)
^{{51}\over{121}}
\exp\biggl(-{{{12\pi^2}\over{11}}{{\beta_I}\over{N^2}}}\biggr).
\label{eqn_Lambda}
\ee
which gives us $\Lambda_I$ in units of a physical mass
scale, $\surd\sigma$, of the SU($N$) gauge theory. 
Using this formula we can extract a value of 
$\surd\sigma/\Lambda_I$ from each of our lattice 
calculations of $a\surd\sigma$. This value will
receive $O(a^2)$ corrections in the usual way,
but it will also receive $O(g^2)$ perturbative
corrections to the two-loop formula in 
eqn(\ref{eqn_Lambda}). These corrections will vary
by very little over our range of $a$ and we would
not be able to determine them numerically. By
contrast the $O(a^2)$ correction varies strongly
so we can perform a continuum extrapolation
\be
{{\surd\sigma(a)} \over {\Lambda^{eff}_I(a)}} =
{{\surd\sigma(0)} \over {\Lambda^{eff}_I(0)}} + c a^2\sigma
\label{eqn_lambdacont}
\ee
where $\Lambda^{eff}_I$ is the value of $\Lambda_I$ renormalised
by an unknown $1+O(\bar{g^2_I})$ factor, where $\bar{g^2_I}$ is 
the (approximately constant) value of  ${g^2_I}$ in our range of
$a$. Performing the continuum extrapolation in 
eqn(\ref{eqn_lambdacont}) we obtain the values shown in
Table \ref{table_lambdacont}. We note that the best continuum 
fit is very poor in the case of SU(2), acceptable for SU(3) 
and good for SU(4) and SU(5). We then find that we obtain
a good fit to these values using just the expected 
leading $O(1/N^2)$ correction:
\be
{{\surd\sigma} \over {\Lambda^{eff}_I}} =
6.05(9) - {{2.65(85)}\over{N^2}}.
\label{eqn_lambdaN}
\ee
That is to say, we find that the $\Lambda$ parameter also
appears to have a smooth limit as $N \to \infty$.
It is perhaps surprising how well this analysis has worked
given the fact that our values of $g_I^2(a)$ are 
not really that small.

\section{Topology and instantons}
\label{sec_topology}

A Euclidean continuum SU($N$) gauge field on a hypertorus
will possess an integer topological charge $Q$. The fluctuations 
of this charge may be characterised by the topological 
susceptibility, $\chi_t \equiv \langle Q^2 \rangle/V$, where
$V$ is the volume of space-time. In QCD one can argue
\cite{eta-thooft}
that these topological fluctuations make the  $\eta^\prime$ 
massive and so solve the axial $U(1)$ problem. In the
large-$N$ limit fermions decouple (for any non-zero quark mass),
so the vacuum topological fluctuations become the same as 
those of the SU($N$) gauge theory. Thus in QCD at large-$N$ and
with $N_f$ flavours, one can relate the mass of the $\eta^\prime$ 
to the topological susceptibility of the  SU($N$) gauge theory
\cite{WV}:
\be
\chi_t \simeq 
{{m^2_{\eta^\prime} f^2_{\eta^\prime}}\over{2N_f}}
\sim (180 \  MeV)^4 .
\label{eqn_WV}
\ee
To obtain the value as $(180 \  MeV)^4$ one assumes that
corrections to this relation are small for SU(3) so that
one can use the experimental values for $m_{\eta^\prime}$
and  $f_{\eta^\prime}$. (One also assumes the latter to be 
equal to $f_\pi$ and one corrects eqn(\ref{eqn_WV}) for
the small pseudoscalar octet contribution.)
There have been many lattice calculations of $ \chi_t $ in
SU(3) intended to test the relation in eqn(\ref{eqn_WV}).
The comparison turns out to be quite successful
(see
\cite{MTrev99}
for a review) but of course it only makes sense if the
finite-$N$ corrections are small for SU(3). We are
not here in a position to estimate the corrections
to the large-$N$ relation $m^2_{\eta^\prime} \propto 1/N$,  
but we can check if the SU(3) topological susceptibility
is close to its large-$N$ limit, and we shall do so
later in this section.

While one hopes that the fluctuations of the topological 
charge are not going to change a great deal when one goes 
from $N = 3$ to  $N = \infty$, the number density of 
instantons, by contrast, is expected to suffer a strong 
exponential suppression 
\cite{WI}:
\be
D(\rho)
\propto 
e^{-{{8\pi^2}\over{g^2(\rho)}}}
=
e^{-{{8\pi^2}\over{\lambda(\rho)}}N}
\label{eqn_WI}
\ee
where $\rho$ is the instanton size and
$\lambda\equiv g^2 N $ is the 't Hooft coupling which 
is kept constant as $N \to \infty$. At first sight it
is hard to see how $\chi_t$ can depend weakly on $N$
if the instanton density depends so strongly.
One can see how this may be 
\cite{MTI}
if we include in our expression for $D(\rho)$ the
factor that arises from the various ways that an
SU(2) subgroup may be embedded in SU($N$):
\be
D(\rho) d\rho
\propto 
{{d\rho}\over{\rho}}{1\over{\rho^4}}
\Biggl\{
{{b^2}\over{\lambda^2(\rho)}}
e^{-{{8\pi^2}\over{\lambda(\rho)}}}
\Biggr\}^N .
\label{eqn_MTI}
\ee
Here $b$ is some known constant and we have also included
the scale-invariant measure, and volume factor. We see
that while at very small $\rho$, where $\lambda(\rho)$ is
small, there will indeed be a strong exponential suppression
of $D(\rho)$ as $N$ increases, this weakens with increasing
$\rho$. At the same time, as $\rho$ increases (anti-)instantons
will begin to overlap and the effective action of an instanton
will begin to decrease. As shown in
\cite{MTI}
the small shift in the effective action calculated in
\cite{CDGI}
is sufficient to reverse the exponential large-$N$ decay,
at values of $\rho$ where such an approximate dilute gas 
calculation remains plausible. Thus we expect
\cite{MTI}
that while for very small $\rho$ there will be a rapid exponential
suppression of $D(\rho)$ with increasing $N$, this will weaken
as $\rho$ increases, and indeed will cease entirely at
some critical but still quite small value of the instanton
size. We shall test this prediction in
Section \ref{subsec_sizeQ}.

We have, in addition, specific expectations about the 
behaviour of $D(\rho)$ as $\rho \to 0$ at fixed $N$.
Here one can do a reliable one-loop perturbative 
calculation around the very small instantons which gives
\be
D(\rho)
\propto 
\rho^{{{11N}\over3}-5} .
\label{eqn_Ismall}
\ee
This will provide a test of our lattice calculations.

\subsection{topology on the lattice}
\label{subsec_latticeQ}

Here we briefly summarise some technical
details of our lattice calculation of the topological
properties of the gauge fields. We use standard
methods which are described and referenced more
fully in 
\cite{MTrev99}.

We start by observing that for smooth fields one can expand 
a plaquette matrix in the $(\mu,\nu)$ plane as
$U_{\mu\nu}(x) = 1 + a^2 F_{\mu\nu}(x) + O(a^4)$.
Thus for smooth fields one can define a lattice
topological charge density $Q_L(x)$ as follows:
\be
Q_L(x) = {1\over{32\pi^2}}
\varepsilon_{\mu\nu\rho\sigma}
Tr\{U_{\mu\nu}(x)U_{\rho\sigma}(x)\}
\stackrel{a \to 0}{\longrightarrow}
a^4 Q(x)
\label{eqn_Qlattice}
\ee
where $Q(x)$ is the continuum topological charge density.

Typical lattice gauge fields are not smooth but are rough
on all scales. So to apply eqn(\ref{eqn_Qlattice}) to
a Monte Carlo generated lattice gauge field we first smoothen
it by a process called `cooling' 
\cite{MTrev99}.
This procedure is just
like the Monte Carlo except that we choose each link matrix
so as to minimise the action (for each Cabibbo-Marinari
SU(2) subgroup). In practice we shall always perform 20 
complete cooling sweeps through the lattice.

Since the cooling algorithm changes only one link matrix at 
a time, the resulting deformation of the fields is local. 
Thus the only way the topological charge can change is by 
an ultraviolet instanton disappearing into a hypercube (or the
reverse). For the physical topological charge to change in
this way, a charge on the scale $\rho \sim 0.5 fm$ must
shrink, under the action of iterated cooling, to the
scale  $\rho \sim a$. (Recall eqn(\ref{eqn_Ismall}) which
tells us that in practice the vacuum contains no very small
instantons.) As $a\to 0$ this will take more and more cooling 
sweeps. In other words, as $a \to 0$ the physical topological
charge becomes stable under our 20 cooling sweeps.

While the total topological charge, $Q$, and hence the
susceptibility $\chi_t = \langle Q^2 \rangle /V$, can
be calculated reliably 
\cite{MTrev99}
in this way, it is much less clear how reliably one
can infer the location and sizes of instantons in 
the vacuum. This is partly because the cooling deforms the 
topological charge density but also because there are
ambiguities in decomposing a distribution of topological 
charge into a `sum' of instantons of various sizes. 
We shall take the most naive approach here. After
the 20 cooling sweeps we assume that any sufficiently
pronounced peak in the $Q_L(x)$ density is due to the
presence of an instanton. We infer the size of the
instanton from the magnitude of the peak using
the classical continuum relation:
\be
Q_{peak} = {6\over{\pi^2\rho^4}}.
\label{eqn_Qpeak}
\ee
In practice we apply a cutoff, and discard any peaks
corresponding to $\rho \geq 10$ (in lattice units). 
In addition if there are two peaks of the same sign
within a distance $2a$ of each other, we keep only the
sharper peak. In this way we try to avoid multiple 
counting of peaks with slightly deformed maxima. This
instanton identification procedure is clearly most reliable
for smaller instantons, since they produce the most prominent
peaks in the topological charge density. However such
instantons will have been deformed by the cooling and
so we should not expect to do better than to identify
some very qualitative features of the instanton
size distribution.

Returning to the total topological charge, we note that
in practice $Q_L = \sum_x Q_L(x)$ need not be very close
to an integer because there will often be topological 
charges which are not very large and these will 
contribute significant $O(a^2/\rho^2)$ corrections to $Q_L$.
Since narrow instantons are easy to identify, and since
the typical value of $Q$ is not large, it is
possible to correct for this. And indeed such corrections
have been applied in many past calculations 
(see for example
\cite{MT88}).
Given our large number of calculations we have chosen not to 
do this by hand (reliable but tedious) but have automated
the process. Our algorithm looks at all the narrow instantons
in the cooled lattice field and depending on their net
charge we shift the real valued topological charge, which
we shall call $Q_r (\equiv Q_L$), to one of the neighbouring
integer values. We call this charge $Q_I$. For the very
coarsest values of $a$ the results differ significantly
from a more careful individual examination (and so we
should expect some disagreement with the older calculations)
but this difference rapidly disappears as $a$ decreases.
In practice this difference will not affect 
our continuum limits in any way. Following previous
calculations
\cite{MT88},
we also calculate
a topological charge, $Q_{Ic}$, which removes from 
$Q_I$ instantons that are sufficiently narrow that they
might be affected by the lattice cut-off. In practice
our criterion for what is narrow is that 
$Q_{peak} \geq 1/16\pi^2$. Using eqn(\ref{eqn_Qpeak})
this corresponds to a cut-off size $\rho \simeq 3$.
Both $Q_r$ and $Q_{Ic}$ are expected to suffer larger
lattice spacing corrections than $Q_{I}$ and so 
we will follow past practice
\cite{MT88}
and use  $Q_{I}$ to obtain our
final continuum values of the topological susceptibility.

In each of our lattice calculations (i.e. given volume and
given $\beta$) we typically calculate the topological charge 
on 2000 lattice gauge fields, spaced by 50 Monte Carlo 
sweeps. This represents an increase by a factor of 5 to 10
in statistics over the older SU(2) and SU(3) calculations
\cite{MTrev99}
that use similar methods.

\subsection{the topological susceptibility}
\label{subsec_susceptibilityQ}

We list our values of $\langle Q^2 \rangle$ in
Tables \ref{table_topsu2} -- \ref{table_topsu5}.
On a lattice of size $L^4$ in lattice units, the
topological susceptibility is 
$a^4\chi_t = \langle Q^2 \rangle/L^4$. Since
we expect $Q = O(V)$, $\chi_t$ should have
a finite limit as $V\to \infty$. It is this finite 
limit we are interested in. To see how large a volume
we need to use in order that finite volume corrections 
should be
negligible we have performed finite volume studies
at $\beta=10.7$ in the case of SU(4), and at 
$\beta=5.90,5.93$ in the case of SU(3). If we convert
the values of $Q_I$ listed in the Tables to
values of $\chi_t$, we see that $\chi_t$ appears
to decrease slightly with increasing volume. Once the 
volume satisfies $aL\surd\sigma \geq 3$ (the case for
the $L=10, 12$ lattices at $\beta=10.7$) any
finite volume effects seem to be within the
$1-2\%$ statistical errors. Thus almost all the
volumes we shall use for our continuum extrapolations
will satisfy this bound.

We see from the Tables that the values of $Q_{Ic}$
rapidly approach the values of $Q_I$ as $a\to 0$,
and that this is more marked as $N$ increases.
This reflects the fact that small instantons are
suppressed and, as we see in eqn(\ref{eqn_Ismall}),
this suppression becomes more severe as $N$ grows.
The values of $Q_r$ also approach  $Q_I$ as $a\to 0$
but here there is less variation with $N$. Having
reassured ourselves that things are as expected, we
shall from now use $Q_I$ as our estimate of $Q$.

\begin	{figure}[p]
\begin	{center}
\epsfig{figure=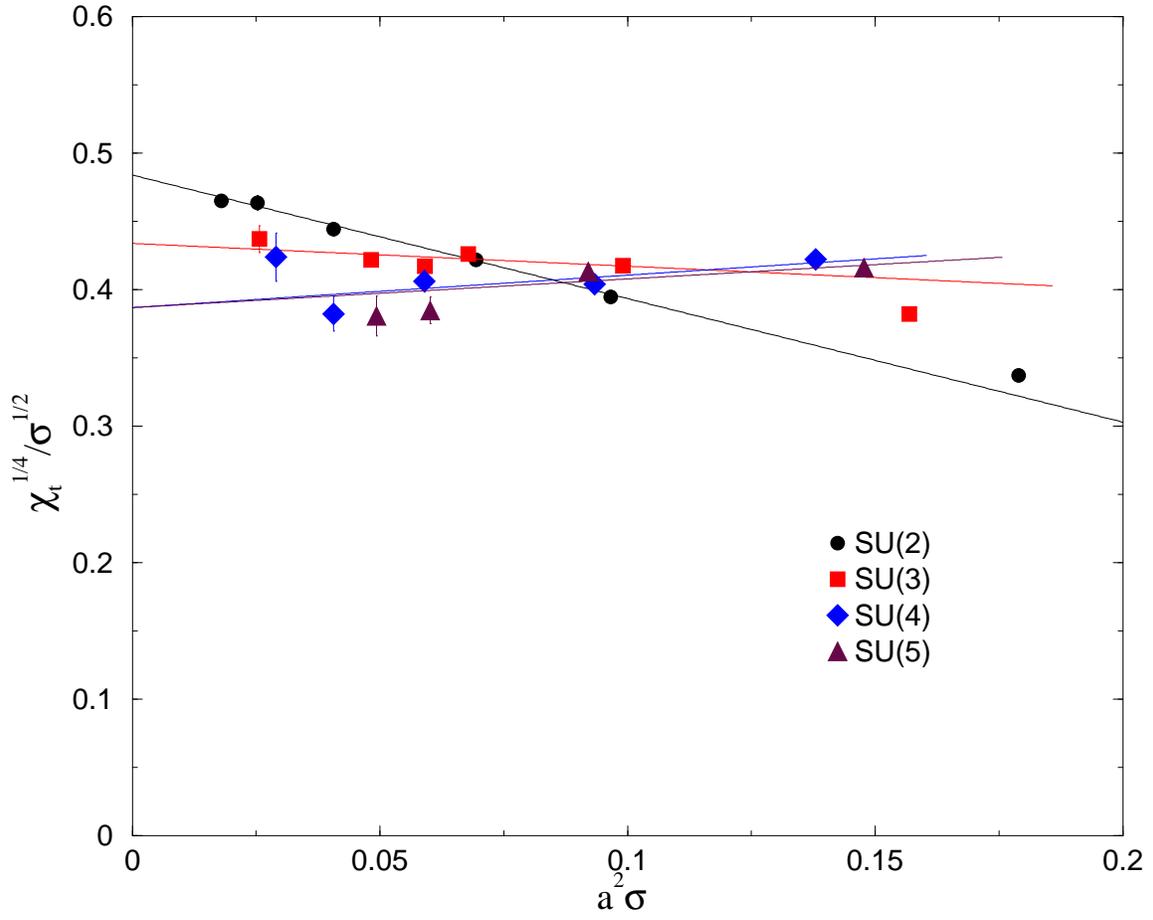, angle=270, width=15cm} 
\end	{center}
\vskip 0.15in
\caption{The topological susceptibility in units of the
string tension plotted against $a^2\sigma$. Continuum
extrapolations for each SU($N$) are shown.}
\label{fig_khi}
\end 	{figure}

In Fig.\ref{fig_khi} we plot the dimensionless ratio
$\chi^{1/4}_t/\surd\sigma$ against $a^2\sigma$.
For sufficiently small $a$ we expect a dependence
\be
{{\chi^{1\over 4}_t(a)} \over {\surd\sigma(a)}} =
{{\chi^{1\over 4}_t(0)} \over {\surd\sigma(0)}} + c a^2\sigma.
\label{eqn_Qcont}
\ee
so that a continuum extrapolation should be a simple
stright line on our plot. We show typical best fits
of this kind.

\begin	{figure}[p]
\begin	{center}
\epsfig{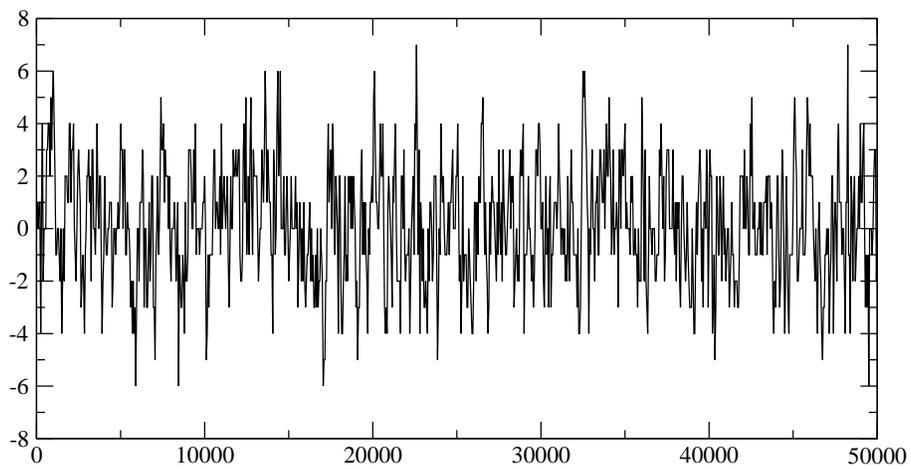} \\
\vspace{3cm}
\epsfig{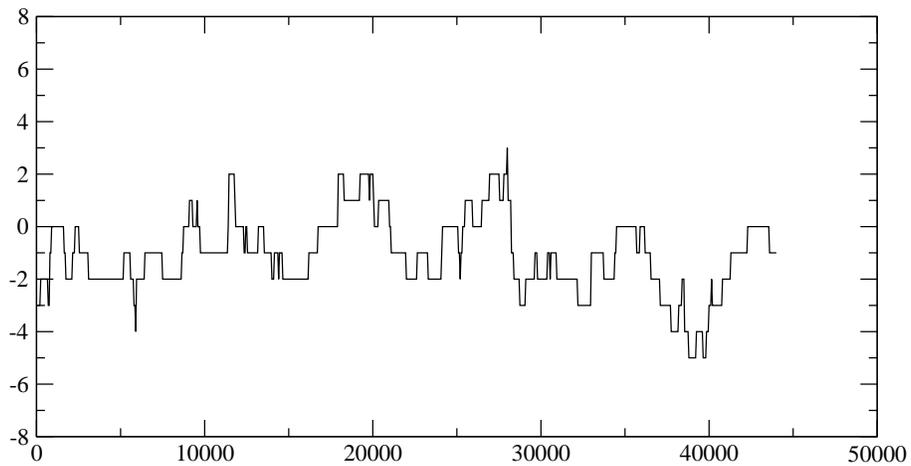}
\end	{center}
\vspace{2cm}
\caption{Two sequences of topological charge calculations, 
made every 50 sweeps, against the number of sweeps. Top
is in SU(3) taken on a $16^4$ lattice at $\beta=6.0$.
Bottom is in SU(5) taken on a $16^4$ lattice at $\beta=17.45$.}
\label{fig_qseq}
\end 	{figure}

We see from Fig.\ref{fig_khi} that the SU(2) and SU(3)
calculations are very precise and under good control.
As we increase $N$ the errors on the values
corresponding to smaller values of $a$ become much
larger, and the increasing scatter of points suggests
that these error estimates are becoming seriously
underestimated. This is to be expected. The Monte
Carlo changes $Q$ by an instanton shrinking through
small values of $\rho$ down to $\rho \sim a$ where
it can vanish through the lattice. (Or the reverse process.) 
However eqn(\ref{eqn_Ismall}) tells us that the probability
of a very small instanton goes rapidly to zero as $N$
grows. Thus the lattice fields rapidly become constrained 
to lie in given topological sectors and for this quantity
the Monte Carlo rapidly ceases to be ergodic as $N$ grows.
This is illustrated in Fig.\ref{fig_qseq}. Here we
show how the value of $Q$ varies in two long Monte Carlo
sequences, the first in SU(3) and the second in SU(5). 
The lattice sizes are the same as is the lattice spacing
(within errors) when expressed in units of the string tension.
Thus the sequences can be directly compared and it is
apparent that the value of $Q$ changes much less frequently
in SU(5) than in SU(3).

\begin	{figure}[p]
\begin	{center}
\epsfig{figure=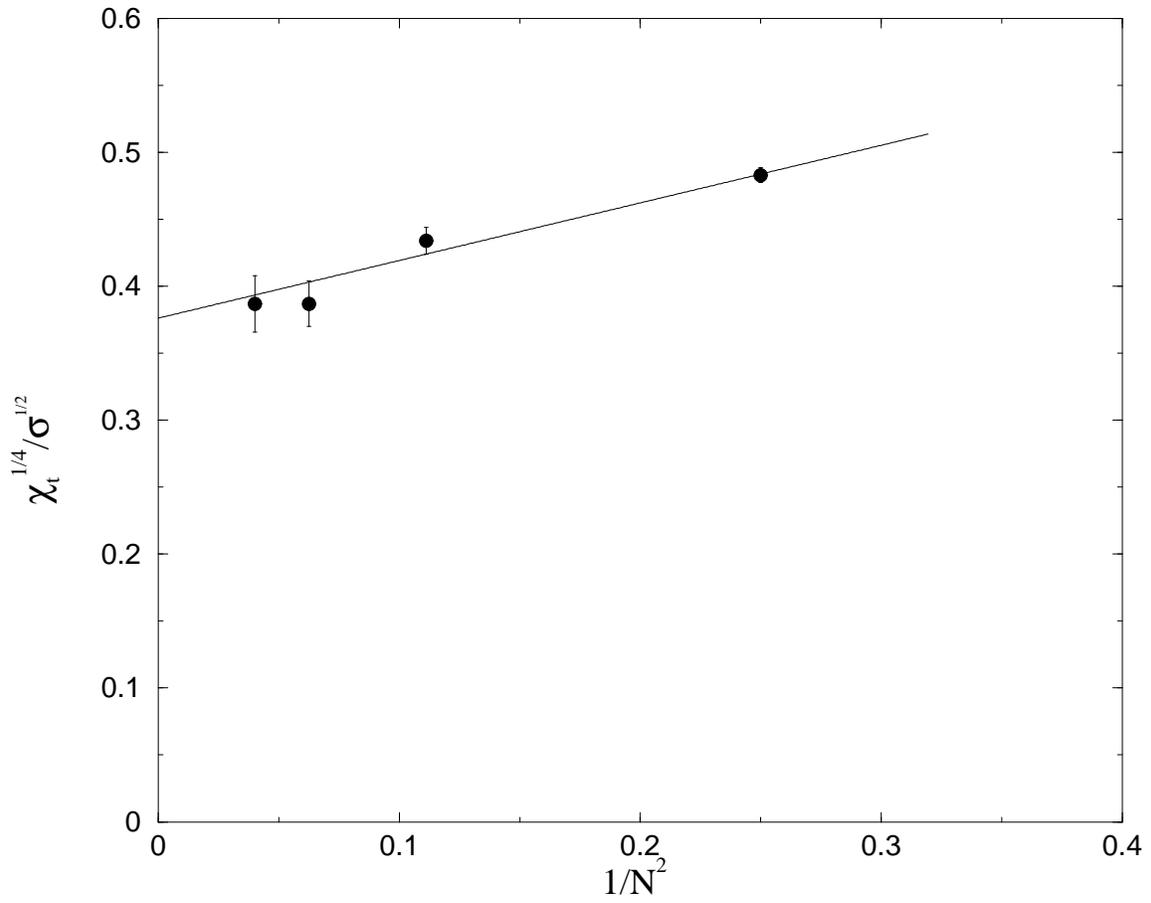, angle=270, width=15cm} 
\end	{center}
\vskip 0.15in
\caption{The continuum topological susceptibility in units 
of the string tension plotted against $1/N^2$. A linear
extrapolation to $N=\infty$ is shown.}
\label{fig_khiN}
\end 	{figure}

In Table \ref{table_continuumQ} we list the continuum 
limit of $\chi^{1/4}_t/\surd\sigma$ for each of our
SU($N$) groups.  In  Fig.\ref{fig_khiN} we plot these
values against $1/N^2$. This is the expected
leading large-$N$ correction, so an extrapolation
to $N=\infty$ would be a straight line on this plot
\be
\left.  {{\chi^{1\over 4}_t} \over {\surd\sigma}} \right |_{N} =
\left.  {{\chi^{1\over 4}_t} \over {\surd\sigma}} \right |_{\infty} 
+ {c \over N^2}.
\label{eqn_QlargeN}
\ee
We show our best such fit in the plot:
\be
  {{\chi^{1\over 4}_t} \over {\surd\sigma}} =
0.376(20)+ {{0.43(10)} \over N^2}.
\label{eqn_QlargeNfit}
\ee
We see from this that the large-$N$ corrections to the 
SU(3) susceptibility are indeed modest. That is to say,
if we express $\lim_{N\to\infty}{\chi_t}$ in physical units
using $\surd\sigma \sim 440 \pm 38 MeV$ 
\cite{MTnewton}
we obtain the value $(165 \pm 17 MeV)^4$ for the large-$N$
susceptibility -- a value that is consistent with our
expectations
\cite{WV}
in eqn(\ref{eqn_WV}).

\subsection{the instanton size distribution}
\label{subsec_sizeQ}

\begin	{figure}[p]
\begin	{center}
\epsfig{figure=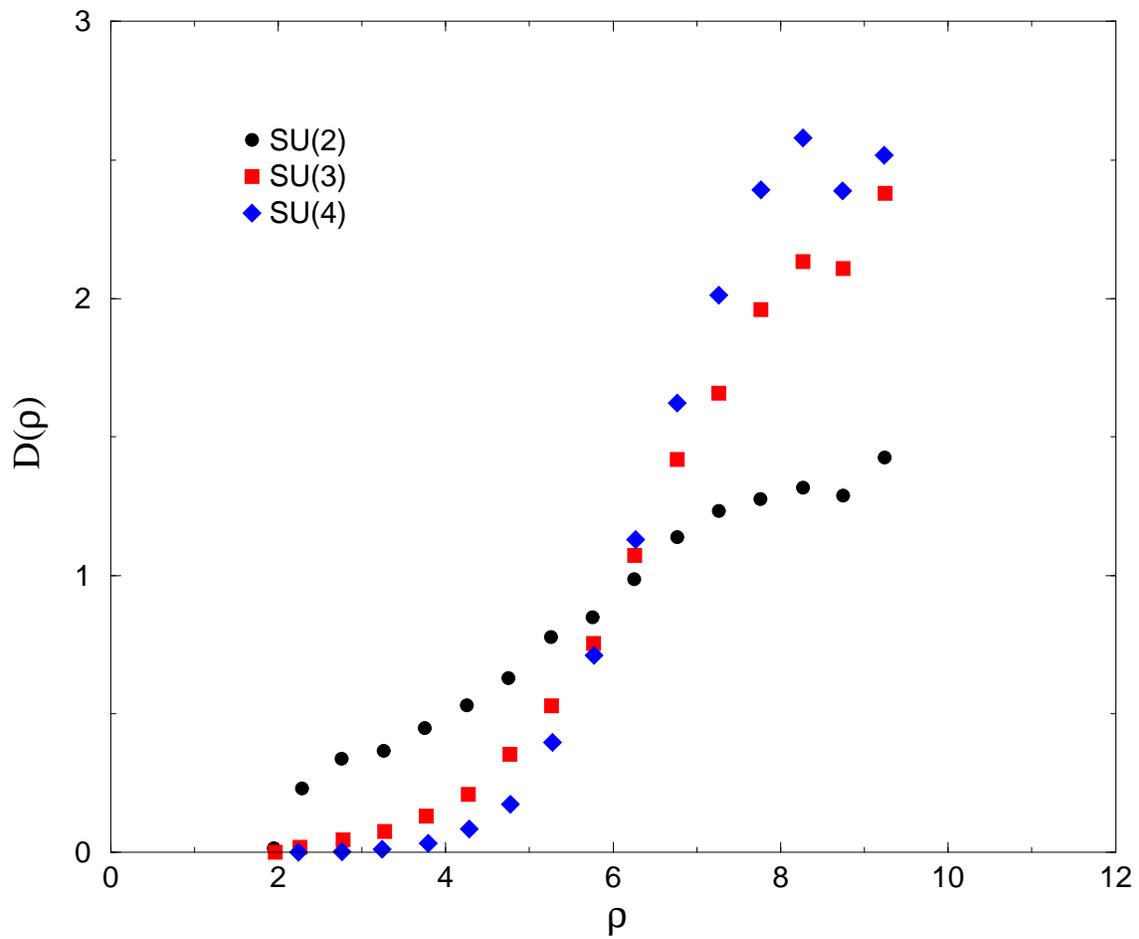, angle=270, width=15cm} 
\end	{center}
\vskip 0.15in
\caption{The number density of topological charges plotted 
as a function of the charge radius, $\rho$; obtained
for $N=2,3,4$ from $20^4$ lattices with 
$a\surd\sigma \sim 0.16$.}
\label{fig_drho}
\end 	{figure}

In Fig.\ref{fig_drho} we plot the number density of 
topological charges against their size $\rho$. We
do this for the $20^4$ SU(2), SU(3) and SU(4) lattices.
This is the calculation with the 
smallest value of $a$ common to these groups.
(In fact the value of $a\surd\sigma$ varies a little 
across these calculations and we have 
rescaled the number density to take this into account.
We have not rescaled the horizontal $\rho$ axis since here
the effect is close to negligible.) 

We observe that the small-$\rho$ tail disappears rapidly
as $N$ increases, just as we expect. This implies, as
we have already discussed, that our Monte Carlo must become
ineffective in exploring different topological charge
sectors as $N$ increases. We observe that the density
$D(\rho)$ appears to tend to a large-$N$ limit where
there are no charges at all with $\rho \leq \rho_c$.
In lattice units $\rho_c \sim 5-6$ which translates
in physical units to $\rho_c \sim 0.9/\surd\sigma \sim 0.4fm$.
The density then rises rapidly and takes non-zero
values over a range of $\rho$. The decrease at small
$\rho$ is very roughly consistent with being
exponential in $N$, as expected.

\begin	{figure}[p]
\begin	{center}
\epsfig{figure=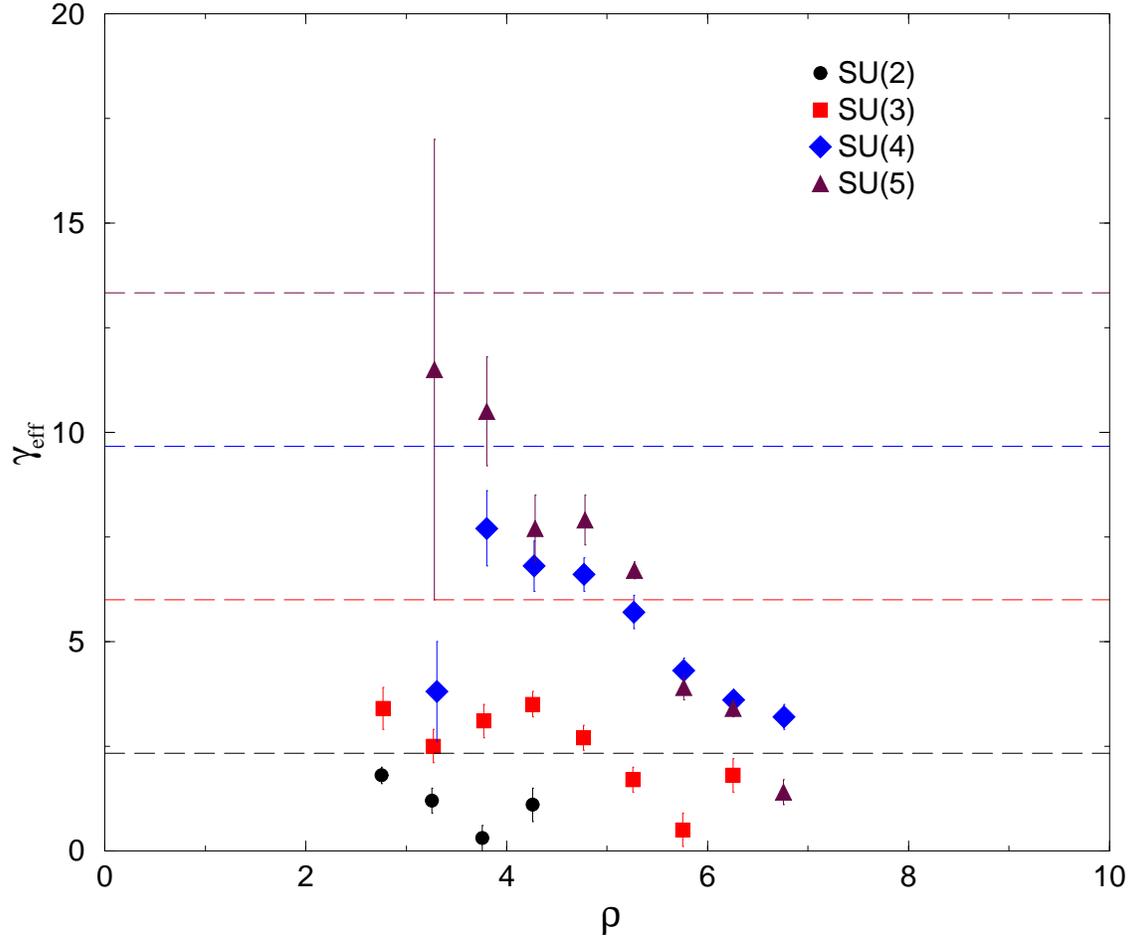, angle=270, width=15cm} 
\end	{center}
\vskip 0.15in
\caption{The effective power dependence of the number density 
of topological charges, $D(\rho) \propto \rho^{\gamma_{eff}}$
compared for $N=2,3,4,5$ on $16^4$ lattices with
$a\surd\sigma \sim 0.21$. Dashed lines are the theoretical
expectations for $\rho \to 0$ taken from eqn(\ref{eqn_Ismall}),
with higher lines correspond to larger $N$.}
\label{fig_Ipower}
\end 	{figure}

As remarked earlier, at small $\rho$ one can make
a reliable theoretical prediction for the 
$\rho$--dependence of $D(\rho)$. To test the prediction
in eqn(\ref{eqn_Ismall}), we fit our calculated
distributions to the form
\be
D(\rho) \propto \rho^{\gamma_{eff}(\rho)}.
\label{eqn_gamma}
\ee
The smallest $a$ common to all our groups is on the
$16^4$ lattices. We extract $D(\rho)$ and $\gamma_{eff}$
from these calculations and plot the resulting  values in 
Fig.\ref{fig_Ipower}. We also plot in each case
the power, from eqn(\ref{eqn_Ismall}), that we expect
to observe for $a \ll \rho \ll 0.5fm$.
Values of $\rho$ satisfying these strong inequalities
do not exist in our calculation. Nonetheless there
is an indication in Fig.\ref{fig_Ipower}
that our values of $\gamma_{eff}$ are indeed tending
to the expected values. This provides evidence
that our number densities, while certainly somewhat
deformed by the cooling, do retain the main features of
the true continuum densities.

\section{Conclusions}
\label{sec_conc}

We have seen, by explicit calculation, that SU($N$)
gauge theories in 3+1 dimensions do indeed appear 
to have a smooth large-$N$ limit that is linearly
confining. We find that, as expected, 
the limit is achieved by keeping
the 't Hooft coupling, $\lambda = g^2N$, fixed.
Remarkably we found that even SU(2) is close to 
SU($\infty$) in the sense that a number of basic mass
ratios can be described over all $N$ using only
a modest $O(1/N^2)$ correction. Thus a number of
simple expressions, such as those in 
eqns(\ref{eqn_scalarN}--\ref{eqn_excscalarN})
and eqn(\ref{eqn_QlargeNfit}), elegantly
encapsulate the corresponding physics for all
values of $N$. Moreover, this greatly increases the
plausibility of arguments that QCD is close to its 
large-$N$ limit.

The purpose of our calculation was not to
obtain the detailed physics of the SU($\infty$)
theory, and for that reason we have not attempted to
compare our results to the predictions of recent
analytic approaches (for example
\cite{SD}
or
\cite{ADS}).
Rather our aim has been to establish whether a detailed
study is, in practice, possible. Clearly it is, and 
in Section \ref{subsec_discuss} we pointed out some
of the lessons we have learned concerning what will
be needed to make such a study successful. Nonetheless, 
our present study has revealed the beginnings
of an interesting $N=\infty$ glueball spectrum, with
$m_{2^{++}} \simeq 1.5 m_{0^{++}}$ and a large
excitation gap in the scalar sector,
$m_{0^{++\star}} \simeq 2 m_{0^{++}}$. However we need 
a much more detailed spectrum if we are to be
able to draw useful dynamical conclusions 
from it. 

We have also seen that the topological 
susceptibility acquires only small  corrections
when  $N$ is reduced from $N=\infty$ to $N=3$. This provides
an {\it a posteriori} justification for estimating
this suceptibility using the experimental
value of $m_{\eta^\prime}$.

As described in the Introduction we shall address the interesting
physics of the new non-trivial $k$-strings that one
encounters for $N\geq 4$ in a parallel publication
\cite{blmtstring2}.
%

%
%
\section*{Acknowledgments}
We thank L. Del Debbio and E. Vicari for interesting discussions.
Our calculations
were carried out on Alpha Compaq workstations in Oxford
Theoretical Physics, funded by PPARC and EPSRC grants.
One of us (BL) thanks PPARC for a postdoctoral fellowship.



%
%
%
%
%

\vfill \eject

\begin{table}
\begin{center}
\begin{tabular}{|c|c|c|c|c|c|c|c|}\hline
\multicolumn{8}{|c|}{SU(2)} \\ \hline
$\beta$ & lattice & sweeps & $am_l$ & $a\surd\sigma$ & $am_{0^{++}}$ & 
$am_{0^{++\star}}$ & $am_{2^{++}}$ \\ \hline
2.25  & $8^4$  & $2\times 10^5$ & 1.301(17) & 0.4231(25) & 1.390(30) 
& 2.5(4)  & 2.30(25)  \\
2.30  & $10^4$ & $10^5$ & 0.861(11) &  0.3108(17) & 1.090(33) 
& 1.58(9) & 1.64(7)  \\
2.40  & $12^4$ & $10^5$ & 0.745(9) &  0.2634(14) & 0.953(19) 
& 1.48(5) & 1.50(5)  \\
2.475 & $16^4$ & $10^5$ & 0.585(8) &  0.2016(13) & 0.754(10) 
& 1.19(2) & 1.111(19) \\
2.55  & $20^4$ & $10^5$ & 0.453(4) &  0.15896(63) & 0.586(10) 
& 0.91(2) & 0.874(15)  \\
2.60  & $24^4$ & $10^5$ & 0.387(4) &  0.13395(62) & 0.514(8) 
& 0.799(10) & 0.750(12) \\ 
\hline
\end{tabular}
\caption{\label{table_datsu2} The mass of the flux loop, $m_l$;
the string tension, $\sigma$, as derived from it using
eqn(\ref{eqn_poly}); the lightest scalar and tensor
glueball masses and the first excited scalar mass. For SU(2) on 
the lattices and at the couplings shown.}
\end{center}
\end{table}

\begin{table}
\begin{center}
\begin{tabular}{|c|c|c|c|c|c|c|c|}\hline
\multicolumn{8}{|c|}{SU(3)} \\ \hline
$\beta$ & lattice & sweeps & $am_l$ & $a\surd\sigma$ & $am_{0^{++}}$ & 
$am_{0^{++\star}}$ & $am_{2^{++}}$  \\ \hline
5.70  & $8^4$  & $10^5$ & 1.124(15) &  0.3961(23) & 0.999(25) 
& 2.07(21) & 2.03(20)  \\
5.80  & $10^4$ & $10^5$ & 0.886(9) &  0.3148(14) & 0.895(28) 
& 1.63(10) & 1.66(6)  \\
5.90  & $10^4$ & $10^5$ & 0.534(9) &  0.2527(18) & 0.736(20) 
& 0.95(7) & 1.22(4)  \\
5.90  & $12^4$ & $10^5$ & 0.727(9) &  0.2605(14) & 0.819(19) 
& 1.40(6) & 1.30(4)  \\
5.93  & $10^3 16$ & $10^5$ & 0.467(7) &  0.2391(15) & 0.708(22) 
& 1.00(7) & 0.94(8)  \\
5.93  & $12^4$ & $10^5$ & 0.6245(62) &  0.2435(11) & 0.756(20) 
& 1.236(36) & 1.21(3)  \\
5.93  & $16^4$ & $10^5$ & 0.879(18) &  0.2430(23) & 0.780(16) 
& 1.487(43) & 1.234(30)  \\
6.00  & $16^4$ & $10^5$ & 0.7071(84) &  0.2197(12) & 0.734(13) 
& 1.03(8) & 1.132(24)  \\
6.20  & $20^4$ & $10^5$ & 0.4603(67) &  0.1601(10) & 0.544(9) 
& 0.955(20) & 0.827(15) \\  \hline
\end{tabular}
\caption{\label{table_datsu3} The mass of the flux loop, $m_l$;
the string tension, $\sigma$, as derived from it using
eqn(\ref{eqn_poly}); the lightest scalar and tensor
glueball masses and the first excited scalar mass. For SU(3) on 
the lattices and at the couplings shown.}
\end{center}
\end{table}

\begin{table}
\begin{center}
\begin{tabular}{|c|c|c|c|c|c|c|c|}\hline
\multicolumn{8}{|c|}{SU(4)} \\ \hline
$\beta$ & lattice & sweeps & $am_l$ & $a\surd\sigma$ & $am_{0^{++}}$ & 
$am_{0^{++\star}}$ & $am_{2^{++}}$   \\ \hline
10.55  & $8^4$  & $2\times 10^5$ & 0.973(17)& 0.3715(28) & 0.837(23) 
& 1.76(11)  & 1.87(10)  \\
10.70  & $6^3 16$ & $0.5\times 10^5$ & 0.268(8) & 0.2716(14) & 0.630(30) 
& 1.44(8) & 0.55(5)  \\
10.70  & $8^3 12$ & $10^5$ & 0.564(10) & 0.2947(21) & 0.780(30) 
& 1.22(4) & 1.14(6)  \\
10.70  & $10^4$ & $10^5$ & 0.8375(92) & 0.3070(15) & 0.906(24) 
& 1.74(10) & 1.53(7)  \\
10.70  & $12^4$ & $10^5$ & 1.033(11) & 0.3055(15) & 0.888(30) 
& 1.64(8) & 1.54(6)  \\
10.90  & $12^4$ & $10^5$ & 0.621(8) & 0.2429(14) & 0.720(30) 
& 1.355(40) & 1.24(4) \\ 
11.10  & $16^4$ & $10^5$ & 0.585(8) & 0.2016(13) & 0.644(17) 
& 1.183(28) & 0.955(55) \\
11.30  & $20^4$ & $10^5$ & 0.5278(65) & 0.1703(10) & 0.572(11) 
& 1.125(20) & 0.885(15)  \\ \hline
\end{tabular}
\caption{\label{table_datsu4} The mass of the flux loop, $m_l$;
the string tension, $\sigma$, as derived from it using
eqn(\ref{eqn_poly}); the lightest scalar and tensor
glueball masses and the first excited scalar mass. For SU(4) on 
the lattices and at the couplings shown.}
\end{center}
\end{table}

\begin{table}
\begin{center}
\begin{tabular}{|c|c|c|c|c|c|c|c|}\hline
\multicolumn{8}{|c|}{SU(5)} \\ \hline
$\beta$ & lattice & sweeps & $am_l$ & $a\surd\sigma$ & $am_{0^{++}}$ & 
$am_{0^{++\star}}$ & $am_{2^{++}}$   \\ \hline
16.755 & $8^4$  & $10^5$ & 1.051(13) & 0.3844(21) & 0.777(23) 
& 1.81(17)  & 2.12(18)  \\
16.975 & $10^4$ & $2\times 10^5$ & 0.816(12) & 0.3034(20) & 0.874(17) 
& 1.63(6) & 1.51(5) \\ 
17.27  & $12^4$ & $1.4\times 10^5$ & 0.634(9) & 0.2452(15) & 0.753(20) 
& 1.39(4) & 1.24(4) \\ 
17.45  & $16^4$ & $10^5$ & 0.724(12) & 0.2221(17) & 0.689(15) 
& 1.15(11) & 1.110(28) \\ \hline 
\end{tabular}
\caption{\label{table_datsu5} The mass of the flux loop, $m_l$;
the string tension, $\sigma$, as derived from it using
eqn(\ref{eqn_poly}); the lightest scalar and tensor
glueball masses and the first excited scalar mass. For SU(5) on 
the lattices and at the couplings shown.}
\end{center}
\end{table}

\begin{table}
\begin{center}
\begin{tabular}{|c||c|c|c|}\hline
\multicolumn{4}{|c|}{continuum limit} \\ \hline
 & $m_{0^{++}}/\surd\sigma$ & $m_{0^{++\star}}/\surd\sigma$
 & $m_{2^{++}}/\surd\sigma$  \\ \hline
SU(2) & 3.844(61) & 6.06(16) & 5.59(15) \\
SU(3) & 3.607(87) & 6.07(30) & 5.13(22) \\
SU(4) & 3.49(14) & 6.80(45) & 5.21(21) \\
SU(5) & 3.38(16) & 6.16(55) & 4.88(38) \\ \hline 
SU($\infty$) & 3.37(15) & 6.43(50) & 4.93(30) \\ \hline 
\end{tabular}
\caption{\label{table_continuumG}
The continuum limit of the lightest scalar and tensor
glueball masses, and the first excited scalar mass,
all in units of the string tension $\sigma$. The
extrapolation of these to $N=\infty$ is also shown.}
\end{center}
\end{table}

\begin{table}
\begin{center}
\begin{tabular}{|c|c|c|c|}\hline
  & $\surd\sigma/\Lambda^{eff}_I$ & range & CL\%  \\ \hline
SU(2) & 5.41(10) &  $\beta\geq 2.475$ & 0.5 \\
SU(3) & 5.70(10) &  $\beta\geq 5.90$  & 17 \\
SU(4) & 5.89(10) &  $\beta\geq 10.70$ & 90 \\
SU(5) & 5.97(8)  &  $\beta\geq 16.75$ & 40 \\ \hline 
\end{tabular}
\caption{\label{table_lambdacont}
The continuum limit of the effective perturbative
$\Lambda_I$ parameter. The range of $\beta$ used and the
confidence level of the best fit is shown.}
\end{center}
\end{table}

\begin{table}
\begin{center}
\begin{tabular}{|c|c|c|c|c|c|}\hline
\multicolumn{6}{|c|}{SU(2)} \\ \hline
$\beta$ & lattice & plaquette & $\langle Q^2_I \rangle$
& $\langle Q^2_{Ic} \rangle$ & $\langle Q^2_r \rangle$ \\ \hline
2.25  & $8^4$  & 0.586207(29) & 1.696(43) & 0.666(19) & 1.397(40) \\ 
2.30  & $10^4$ & 0.616955(29) & 2.268(71) & 1.159(46) & 1.922(60) \\
2.40  & $12^4$ & 0.629995(17) & 3.158(86) & 1.664(56) & 2.812(86) \\ 
2.475 & $16^4$ & 0.646921(10) & 4.23(15) & 2.97(10) & 3.84(14) \\ 
2.55  & $20^4$ & 0.661367(4) & 4.73(19) & 3.53(12) & 4.30(18) \\
2.60  & $24^4$ & 0.670009(3) & 5.00(15) & 4.22(14) & 4.55(14) \\ \hline
\end{tabular}
\caption{\label{table_topsu2}
The fluctuation of the topological charge using three
different measures of the charge (see Section \ref{subsec_latticeQ}). 
Also the average plaquette. For SU(2).} 
\end{center}
\end{table}

\begin{table}
\begin{center}
\begin{tabular}{|c|c|c|c|c|c|}\hline
\multicolumn{6}{|c|}{SU(3)} \\ \hline
$\beta$ & lattice &  plaquette & $\langle Q^2_I \rangle$
& $\langle Q^2_{Ic} \rangle$ & $\langle Q^2_r \rangle$ \\ \hline
5.70  & $8^4$  & 0.549123(56) & 2.151(60) & 1.002(38) & 1.779(59) \\
5.80  & $10^4$ & 0.567633(12) & 2.986(68) & 1.767(45) & 2.547(63) \\
5.90  & $10^4$ & 0.58187(3) & 1.452(62) & 1.042(38) & 1.205(48) \\
5.90  & $12^4$ & 0.58185(2) & 3.147(98) & 2.321(75) & 2.675(91) \\
5.93  & $10^3 16$  & 0.58560(3) & 1.830(63) & 1.415(53) & 1.527(53) \\ 
5.93  & $12^4$ & 0.585600(15) & 2.655(90) & 2.066(68) & 2.266(78) \\ 
5.93  & $16^4$ & 0.585580(10) & 6.92(25) & 5.49(22) & 6.47(24) \\ 
6.00  & $16^4$ & 0.593669(8) & 4.83(17) & 4.12(14) & 4.30(16) \\ 
6.20  & $20^4$ & 0.613622(7) & 3.85(33) & 3.72(33) & 3.49(31) \\  \hline
\end{tabular}
\caption{\label{table_topsu3}
The fluctuation of the topological charge using three
different measures of the charge (see Section \ref{subsec_latticeQ}). 
Also the average plaquette. For SU(3).} 
\end{center}
\end{table}

\begin{table}
\begin{center}
\begin{tabular}{|c|c|c|c|c|c|}\hline
\multicolumn{6}{|c|}{SU(4)} \\ \hline
$\beta$ & lattice &  plaquette & $\langle Q^2_I \rangle$
& $\langle Q^2_{Ic} \rangle$ & $\langle Q^2_r \rangle$ \\ \hline
10.55  & $8^4$  & 0.537290(50) & 2.48(8) & 1.29(5) & 2.05(7) \\
10.70  & $8^3 12$ & 0.554114(31) & 1.477(54) & 1.114(35) & 1.213(43) \\
10.70  & $10^4$ & 0.554100(20) & 2.56(9) & 1.99(8) & 2.15(8) \\
10.70  & $12^4$ & 0.554105(14) & 4.82(16) & 3.84(13) & 4.25(15) \\
10.90  & $12^4$ & 0.570103(10) & 1.97(14) & 1.82(12) & 1.69(11) \\
11.10  & $16^4$ & 0.583332(5) & 2.31(32) & 2.26(32) & 2.08(29) \\
11.30  & $20^4$ & 0.595014(4) & 4.34(76) & 4.31(74) & 3.98(70) \\  \hline
\end{tabular}
\caption{\label{table_topsu4}
The fluctuation of the topological charge using three
different measures of the charge (see Section \ref{subsec_latticeQ}). 
Also the average plaquette. For SU(4).} 
\end{center}
\end{table}

\begin{table}
\begin{center}
\begin{tabular}{|c|c|c|c|c|c|}\hline
\multicolumn{6}{|c|}{SU(5)} \\ \hline
$\beta$ & lattice &  plaquette & $\langle Q^2_I \rangle$
& $\langle Q^2_{Ic} \rangle$ & $\langle Q^2_r \rangle$ \\ \hline
16.755 & $8^4$  & 0.52783(5) & 2.684(68) & 1.674(51) & 2.258(67) \\
16.975 & $10^4$ & 0.545164(14) & 2.470(75) & 2.222(63) & 2.083(67) \\
17.27  & $12^4$ & 0.561139(7) & 1.65(16) & 1.61(16) & 1.44(14) \\
17.45  & $16^4$ & 0.569407(5) & 3.36(53) & 3.35(52) & 3.00(47) \\ \hline
\end{tabular}
\caption{\label{table_topsu5}
The fluctuation of the topological charge using three
different measures of the charge (see Section \ref{subsec_latticeQ}). 
Also the average plaquette. For SU(5).} 
\end{center}
\end{table}

\begin{table}
\begin{center}
\begin{tabular}{|c||c|c|c|}\hline
\multicolumn{4}{|c|}{$\chi^{1/4}_t/\surd\sigma$ : continuum limit} \\ \hline
 & $Q_I$ & $Q_{Ic}$ & $Q_r$ \\ \hline
SU(2) & 0.4831(56) & 0.4745(63) & 0.4742(56) \\
SU(3) & 0.434(10)  & 0.451(8)   & 0.427(11)  \\
SU(4) & 0.387(17)  & 0.426(15)  & 0.380(16)  \\ 
SU(5) & 0.387(21)  & 0.350(30)  & 0.374(20)  \\ \hline 
\end{tabular}
\caption{\label{table_continuumQ}
The continuum limit of the topological susceptibility $\chi_t$
in units of the string tension $\sigma$, for the three
different measures of the topological charge indicated.}
\end{center}
\end{table}

\clearpage

\end{document}